\documentclass[a4paper,twoside,11pt]{article} % to use for all kind of papers but AACA
\newlength{\defaultparindent}
\usepackage{latexsym}
\usepackage{amsmath}
\usepackage{amsfonts}
\usepackage{amsthm}
\usepackage{bbold}
\usepackage[applemac]{inputenc} % Così le "è" vengono trasformate in \`e etc. etc.
\usepackage{multirow}
\usepackage{hyperref}
\hypersetup{colorlinks=true,citecolor=green,linkcolor= blue}
\usepackage{todonotes} % more appealing than \marginpar ??

\usepackage[arXiv]{optional} % Possible values of the option fixing the format: x, std, arXiv, JMP, JOPA, AACA and also note control: margin_notes, final_notes
\def\mynote{\todo} % \marginpar or \todo
\opt{AACA}{% only for submission to AACA -> \cal font is not defined....
\def\cal{\mathcal}
}

\opt{x,std,AACA}{% in all cases - but arXiv, JMP & JOPA - include my file of standard defs: mbh_defs_TeX
\input{mbh_defs_TeX}% My standard TeX definitions, original in Dictionaries & Gloassaries ƒ
}

\opt{arXiv,JMP,JOPA}{% only for arXiv, JMP & JOPA we need the definitions included in mbh_defs_TeX (that contains original versions)

\newtheorem{MS_theorem}{Theorem}
\newtheorem{MS_Proposition}{Proposition}
\newtheorem{MS_Corollary}[MS_Proposition]{Corollary}
\def\myconjugate#1{\bar{#1}} % this way conjugate definitions can be easily changed
\def\ie{i.e.\ }
\def\eg{e.g.\ }
\def\End{\textrm{End}}
\newcommand{\R}{\ensuremath{\mathbb{R}}} % good; MB XII 2005; needs \usepackage{bbold}
\newcommand{\C}{\ensuremath{\mathbb{C}}} % good; MB XII 2005; needs \usepackage{bbold}
\newcommand{\F}{\ensuremath{\mathbb{F}}} % good; MB XII 2005; needs \usepackage{bbold}
\newcommand{\Identity}{\ensuremath{\mathbb{1}}} % good; MB XII 2005; needs \usepackage{bbold}
\def\my_span#1{\mbox{Span}\left(#1\right)} % changed from 'span' since it interfered with \multicolumn{} MB X 2009
\def\mytrace#1{\mbox{Tr}\left(#1\right)} % if I call this my_trace it messes with my_span ??? MB VII 2012
\def\qed{$\Box$}
\def\bino#1#2{\left( \! \begin{array}{c} #1 \\ #2 \end{array} \! \right)}
\def\dotinformula{\;\; \mathrm{.}} % defines space + a full stop (in \rm font) to be placed at the end of a formula
\def\Pin#1{\ensuremath{\mbox{Pin}\left(#1\right)}}

\newcommand{\JJ}{\mathbin{\raisebox{0.25ex}{$\scriptstyle % originally was \footnotesize but is "invalid in math mode" - mbh
                       \rm\vphantom{I}%
                       \_\hskip -0.25em\_%
                       \vrule width 0.6pt$}}}           %left contraction (from Makro98.tex)
\newcommand{\comm}[2]{\ensuremath{\left[ #1, #2 \right]}}
\newcommand{\anticomm}[2]{\ensuremath{\left\{ #1, #2 \right\}}} % or \left[...\right]^+
\newcommand{\myCl}[3]{\ensuremath{{{\cal C}\ell}_{#1, #2} {\left( #3 \right)}}}

}

\opt{JMP}{% only for submission to JMP -> double line spacing
}

\def\h_eigen{\eta}
\def\g_eigen{\theta}

\begin{document}

\opt{x,std,arXiv,JMP,JOPA}{% in all cases - but AACA
\title{{\bf On Spinors and Null Vectors} %\\(temporary title)
	}

\author{\\
	\bf{Marco Budinich}%
\footnote{on leave of absence from: University of Trieste, Trieste, Italy}%
\\
	ICTP and INFN, Trieste, Italy\\
	\texttt{mbh@ts.infn.it}\\
%	\texttt{http://www.ts.infn.it/\~{ }mbh/MBHgeneral.html}\\
%
%	Very preliminary - restricted circulation (FYEO)
%
%
%	Submitted
%	Submitted to: {\em Journal of Mathematical Physics} on August 3, 2012
%\\	Submitted to: {\em Communications in Mathematical Physics} on May 10, 2013.\\
\\	Published in: {\em Journal of Physics A: Math. and Th.}\\
%(submitted on July 6, 2013 and in amended version{\bf{s}} on Sep. 8 and Dec. 16)\\
%	Submitted to: {\em Journal of Physics A: Mathematical and Theoretical} \\on December 21, 2010
%	Submitted to: {\em Letters in Mathematical Physics} on July 28, 2011
%	Submitted to: {\em Advances in Applied Clifford Algebras} on September 5, 2011
%	Published in: {\em Advances in Applied Clifford Algebras}, 2011\\
%	{\small DOI:10.1007/s00006-011-0316-2}
	}
%\date{ \today }
\date{ } % to hide the date this line must be present (believe it or not...)
\maketitle
}

\opt{AACA}{% only for AACA
\title[On Spinors and Null Vectors]{On Spinors and Null Vectors}

\author{Marco Budinich}
\address{Dipartimento di Fisica\\
	Università di Trieste \& INFN\\
	Via Valerio 2, I - 34127 Trieste, Italy}
\email{mbh@ts.infn.it}
}

\begin{abstract}
We investigate the relations between spinors and null vectors in Clifford algebra of any dimension with particular emphasis on the conditions that a spinor must satisfy to be simple (also: pure). In particular we prove: i) a new property for null vectors: each of them bisects spinor space into two subspaces of equal size; ii) that simple spinors form one-dimensional subspaces of spinor space; iii) a necessary and sufficient condition for a spinor to be simple that generalizes a theorem of Cartan and Chevalley which becomes a corollary of this result. We also show how to write down easily the most general spinor with a given associated totally null plane.
\end{abstract}

%\opt{x,std,arXiv,JMP,JOPA}{% in all cases - but AACA
%\noindent{\bf Keywords:} {Clifford algebra, simple spinors, mathematical physics, Fock basis}
%}

\opt{AACA}{% only for AACA
\maketitle
}

\section{Introduction}
\label{Intro}
Exactly a century ago {\'{E}}lie Cartan \cite{Cartan_1913, Cartan_1937} introduced spinors that were later thoroughly investigated by Claude Chevalley \cite{Chevalley_1954} in the mathematical frame of Clifford algebra; in this work spinors were identified as elements of minimal left ideals of the algebra. The interplay between spinors and null (also: isotropic) vectors, pioneered by Cartan, and thus sometimes called the Cartan map, is central and have been visited many times since then, see \eg \cite{BudinichP_1989, Keller_1991} and references therein.
\opt{margin_notes}{\mynote{mbh.note: see also Paolo's papers}}
This relation is pivotal to many fields of physics, the Weyl equation being just one prominent application.

Among spinors, simple (also: pure) spinors play a principal role both in this relation and in many fields in physics like string theory, gravity and supergravity and also in geometry \cite{Berkovits_2000_0, Berkovits_Nekrasov_2005, Gualtieri_2003} and the characterization of simple spinors is thus relevant for many applications.
\opt{margin_notes}{\mynote{mbh.note: possible citation http://dx.doi.org/10.1007/s00220-009-0881-6}}

Finding properties to identify simple spinors has proved to be an elusive subject and the main available result is a theorem due to Cartan and Chevalley (see \eg \cite{Trautman_Trautman_1994} proposition~5) stating that a spinor is simple iff a certain number of constraints are satisfied. Unfortunately, the number of constraints grows exponentially with the dimension of the vector space that render its use impractical already in spaces of moderate dimension. Up to now this has been the only available result to characterize simple spinors and is about $60$ years old indicating that the subject is mature, which is not to say that everybody is familiar with it.

In this paper we address the relation between spinors and null vectors and will present two different means of characterizing simple spinors. Simple spinors are known to be in one to one correspondence with vector subspaces of null vectors and of maximal dimension. We will exploit this property to show that simple spinors correspond to one-dimensional subspaces of spinor space and this will allow us to write down immediately the most general simple spinor corresponding to a given, maximal, totally null subspace. Afterwards, we will prove a necessary and sufficient condition for a spinor to be simple that includes previous results and in particular the quoted theorem of Cartan and Chevalley that will appear as a particular case of this more general result.

We will investigate relations between spinors and null vectors in $\C^{2 m}$ and $\R^{2 m}$ with signature $(m,m)$, a standard choice in these studies,
\opt{margin_notes}{\mynote{mbh.note: and twistor.... see Wiki + Witten ?}}
exploiting the Extended Fock Basis (EFB) of Clifford algebra \cite{BudinichM_2009, Budinich_2011_EFB}, recalled in section~\ref{Clifford_algebra_and_EFB}. With this basis \emph{any} element of the algebra can be expressed in terms of simple spinors: from scalars to vectors and multivectors. Sections~\ref{Vector_space_V} and \ref{Spinor_spaces} are dedicated, respectively, to the vector space $V$ and to the spinor space(s) $S$ of Clifford algebra. In this last section we show how one can concisely represent the most general spinor corresponding to a given vector subspace made entirely of null vectors.
\opt{margin_notes}{\mynote{mbh.note: Dirac linear structure ? see Gualtieri PhD thesis; T. Courant, Dirac structures, Trans. A.M.S. 319 (1990), 631-661.}}

Section~\ref{Simple_spinors} deals with simple spinors and conveys the main result: a necessary and sufficient condition for a spinor to be simple.

For the convenience of the reader we tried to make this paper as elementary and self-contained as possible.

\section{The extended Fock basis of Clifford algebra}
\label{Clifford_algebra_and_EFB}
We start summarizing the essential properties of the EFB introduced in \cite{BudinichM_2009} and \cite{Budinich_2011_EFB}. We consider Clifford algebras \cite{Chevalley_1954} over field $\F$, with an even number of generators $\gamma_1, \gamma_2, \ldots, \gamma_{2 m}$, a vector space $\F^{2 m} := V$ and a scalar product $g$: these are simple, central, algebras of dimension $2^{2 m}$. As usual
$$
2 g(\gamma_i, \gamma_j) = \gamma_i \gamma_j + \gamma_j \gamma_i := \anticomm{\gamma_i}{\gamma_j}
$$
and we stick to $\F = \R$ with signature $V = \R^{m, m}$; $g(\gamma_i, \gamma_j) = \delta_{i j} (-1)^{i+1}$ \ie
\begin{equation}
\label{space_signature}
\left\{ \begin{array}{l l l}
\gamma_{2 i - 1}^2 & = & 1 \\
\gamma_{2 i}^2 & = & -1
\end{array} \right.
\qquad i = 1,\ldots,m
\end{equation}
but results also hold for $\F = \C$. Given the $\R^{m, m}$ signature we indicate the Clifford algebra with \myCl{m}{m}{g}.

A Clifford algebra is the direct sum of its graded parts: field $\F := \F^{(0)}$, vectors $V := \F^{(1)}$ and multivectors $\F^{(k)}, \; 1 < k \le 2 m$
\begin{equation}
\label{Clifford_direct_sum}
\myCl{m}{m}{g} = \F^{(0)} \oplus \F^{(1)} \oplus \cdots \oplus \F^{(2 m)}
\end{equation}
\opt{margin_notes}{\mynote{mbh.note: referee \#$2$ blamed on ``graded isomorphic'', see a discussion at p. $506'$}}
and is isomorphic to $\F( 2^m )$, the algebra of matrices of size $2^m \times 2^m$.

The Witt, or null, basis of the vector space $V$ is defined:
\begin{equation}
\label{formula_Witt_basis}
\left\{ \begin{array}{l l l}
p_{i} & = & \frac{1}{2} \left( \gamma_{2i-1} + \gamma_{2i} \right) \\
q_{i} & = & \frac{1}{2} \left( \gamma_{2i-1} - \gamma_{2i} \right)
\end{array} \right.
\Rightarrow
\left\{\begin{array}{l l l}
\gamma_{2i-1} & = & p_{i} + q_{i} \\
\gamma_{2i} & = & p_{i} - q_{i}
\end{array} \right.
\quad i = 1,2, \ldots, m
\end{equation}
that, with $\gamma_{i} \gamma_{j} = - \gamma_{j} \gamma_{i}$, easily gives
\begin{equation}
\label{formula_Witt_basis_properties}
\anticomm{p_{i}}{p_{j}} = \anticomm{q_{i}}{q_{j}} = 0
\qquad
\anticomm{p_{i}}{q_{j}} = \delta_{i j}
\end{equation}
that imply $p_i^2 = q_i^2 = 0$, at the origin of the name ``\emph{null}'' given to these vectors.

Following Chevalley we define spinors as elements of a minimal left ideal we will indicate with $S$%
\footnote{in an algebra $A$ a subset $S$ is a left ideal if for any $a \in A, \varphi \in S \implies a \varphi \in S$; it is minimal if it does not contain properly any other ideal. For example in matrix algebra the subset of matrices with only one nonzero column form a minimal left ideal.}%
. Simple spinors are those elements of $S$ that are annihilated by a null subspace of $V$ of maximal dimension.

\bigskip

The EFB of \myCl{m}{m}{g} is given by the $2^{2 m}$ different sequences
\begin{equation}
\label{EFB_def}
\psi_1 \psi_2 \cdots \psi_m := \Psi \qquad \psi_i \in \{ q_i p_i, p_i q_i, p_i, q_i \} \qquad i = 1,\ldots,m
\end{equation}
in which each $\psi_i$ is either a vector or a bi--vector and we will reserve $\Psi$ for EFB elements. The main characteristics of EFB is that all its elements are \emph{simple} spinors \cite{BudinichM_2009, Budinich_2011_EFB}.

The EFB essentially extends to the entire algebra the Fock basis~\cite{BudinichP_1989} of its spinor spaces and, making explicit the construction $\myCl{m}{m}{g} \cong \overset{m}{\otimes} \myCl{1}{1}{g}$, allows one to prove in \myCl{1}{1}{g} many properties of \myCl{m}{m}{g}%
\opt{margin_notes}{\mynote{mbh.note: identity: $\anticomm{p_{i}}{q_{j}} = \delta_{i j} \Identity$ or $\anticomm{p_{i}}{q_{j}} = \delta_{i j}$ ? Think: $\mytrace{\gamma_i^2} = 2^m$ while $\mytrace{p_i q_i} = 2^{m-1}$; see p. 394.3}}
\footnote{A technical remark: whereas it is customary to see Clifford algebra as a direct sum of its graded parts (\ref{Clifford_direct_sum}), these parts are no more evident in EFB where all elements are multivectors with grade between $m$ and $2m$. Consequently whereas the notation $\gamma_{2 i - 1}^2 = 1$ is imprecise but usually acceptable, in EFB (\ref{formula_Witt_basis_properties}) appears harder to digest since in EFB there are no field elements.

In EFB $\Identity = \anticomm{q_1}{p_1} \anticomm{q_2}{p_2} \cdots \anticomm{q_m}{p_m}$ that agrees with $\mytrace{\Identity} = 2^m = \mytrace{\gamma_{2 i - 1}^2}$ and in EFB $\mytrace{\anticomm{q_{i}}{p_{i}}} = \mytrace{\anticomm{q_1}{p_1} \anticomm{q_2}{p_2} \cdots \anticomm{q_m}{p_m}} = 2^m$. On the other hand $\mytrace{p_{i}q_{i}} = 2^{m-1}$ and the trace of one of the $2^m$ EFB elements forming the expansion of $\anticomm{q_1}{p_1} \anticomm{q_2}{p_2} \cdots \anticomm{q_m}{p_m}$ has $\mytrace{\Psi} = 1$ and they represent primitive idempotents. All in all we will accept to trade rigor for clarity and we will omit the identity symbol $\Identity$ where it would be formally needed and also omit unnecessary terms and write $\anticomm{p_{i}}{q_{i}} = 1$.
}%
.

\subsection{$h-$ and $g-$signatures}
\label{hg_signatures}
We start observing that $\gamma_{2 i - 1} \gamma_{2 i} = q_i p_i - p_i q_i := \comm{q_i}{p_i}$ and that for $i \ne j$ $\comm{q_i}{p_i} \psi_j = \psi_j \comm{q_i}{p_i}$. With (\ref{formula_Witt_basis_properties}) and (\ref{EFB_def}) it is easy to calculate
\begin{equation}
\label{commutator_property}
\comm{q_i}{p_i} \psi_i = h_i \psi_i \qquad h_i = \left\{
\begin{array}{l l}
+1 & \quad \mbox{iff $\psi_i = q_i p_i \; \mbox{or} \; q_i$} \\
-1 & \quad \mbox{iff $\psi_i = p_i q_i \; \mbox{or} \; p_i$}
\end{array} \right.
\end{equation}
and the value of $h_i$ depends on the first null vector appearing in $\psi_i$. We have thus proved that $\comm{q_i}{p_i} \Psi = h_i \Psi$. In EFB the identity $\Identity$ and the volume element $\Gamma$ have similar expressions:
\begin{eqnarray*}
\Identity & := & \anticomm{q_1}{p_1} \anticomm{q_2}{p_2} \cdots \anticomm{q_m}{p_m} \\
\Gamma & := & \gamma_1 \gamma_2 \cdots \gamma_{2 m} = \comm{q_1}{p_1} \comm{q_2}{p_2} \cdots \comm{q_m}{p_m}
\end{eqnarray*}
with which
\begin{equation}
\label{Weyl_eigenvector}
\Gamma \Psi = \h_eigen \; \Psi \qquad \h_eigen := \prod_{i = 1}^m h_i = \pm 1 \dotinformula
\end{equation}
Each EFB element $\Psi$ has thus an ``$h-$signature'' that is a vector $(h_1, h_2, \ldots, h_m) \in \{ \pm 1 \}^m$ and the eigenvalue $\h_eigen$ is the {\em chirality}. Similarly, the ``$g-$signature'' of an EFB element is the vector $(g_1, g_2, \ldots, g_m) \in \{ \pm 1 \}^m$ where $g_i$ is the parity of $\psi_i$ under the main algebra automorphism $\gamma_i \rightarrow - \gamma_i$. With this definition and with (\ref{commutator_property}) we can easily derive that
\begin{equation}
\label{commutator_left_property}
\psi_i \comm{q_i}{p_i} = g_i \comm{q_i}{p_i} \psi_i = h_i g_i \psi_i
\end{equation}
and thus
\begin{equation}
\label{EFB_are_left_Weyl}
\Psi \; \Gamma = \h_eigen \g_eigen \; \Psi \qquad \h_eigen \g_eigen = \pm 1 \qquad \g_eigen := \prod_{i = 1}^m g_i
\end{equation}
where the eigenvalue $\h_eigen \g_eigen$ is the product of chirality times $\g_eigen$, the global parity of the EFB element $\Psi$ under the main algebra automorphism. We can resume saying that all EFB elements are not only Weyl eigenvectors, \ie right eigenvectors of $\Gamma$ (\ref{Weyl_eigenvector}), but also its left eigenvectors (\ref{EFB_are_left_Weyl}) with respective eigenvalues $\h_eigen$ and $\h_eigen \g_eigen$.

\subsection{EFB formalism}
\label{EFB_formalism}
$h-$ and $g-$signatures play a crucial role in this description of \myCl{m}{m}{g}: first of all one easily sees that any EFB element $\Psi = \psi_1 \psi_2 \cdots \psi_m$ is uniquely identified by its $h-$ and $g-$signatures: $h_i$ determines the first null vector ($q_i$ or $p_i$) appearing in $\psi_i$ and $g_i$ determines if $\psi_i$ is even or odd.

It can be shown \cite{Budinich_2011_EFB} that \myCl{m}{m}{g}, as a vector space, is the direct sum of its $2^m$ subspaces of:
\begin{itemize}
\item different $h-$signatures or:
\item different $g-$signatures or:
\item different $h \circ g-$signatures (where $h \circ g$ is the Hadamard (entrywise) product of $h-$ and $g-$signatures vectors).
\end{itemize}

We can thus uniquely identify each of the $2^{2 m}$ EFB elements with any two of these three ``indices''. Since different $h \circ g-$signatures will identify different spinor spaces, denoted $S_{h \circ g}$, it is convenient to choose respectively the $h-$signature and the $h \circ g-$signature \ie
\begin{equation*}
\Psi_{a b} \left\{
\begin{array}{l l}
a \in \{ \pm 1 \}^m \quad \mbox{is the} \quad h-\mbox{signature} \\
b \in \{ \pm 1 \}^m \quad \mbox{is the} \quad h \circ g-\mbox{signature}
\end{array} \right.
\end{equation*}
so that the generic element of $\mu \in \myCl{m}{m}{g}$ can be written as $\mu = \sum_{a b} \xi_{a b} \Psi_{a b}$ with $\xi_{a b} \in \F$. With this choice of the indices one can prove \cite{Budinich_2011_EFB} that:
\begin{equation}
\label{EFB_products}
\Psi_{a b} \Psi_{c d} = s(a,b,d) \, \delta_{b c} \Psi_{a d} \qquad s(a,b,d) = \pm 1
\end{equation}
where $\delta_{b c}$ is $1$ if and only if the two signatures $b$ and $c$ are equal and the sign $s(a,b,d)$, quite tedious to calculate, depends on the indices; in \cite{Budinich_2011_EFB} it is shown how it can be calculated with matrix isomorphism. With this result one can calculate the generic Clifford product
\begin{eqnarray*}
\mu \nu & = & \left(\sum_{a b} \xi_{a b} \Psi_{a b} \right) \left(\sum_{c d} \zeta_{c d} \Psi_{c d} \right) = \sum_{a b c d} \xi_{a b} \zeta_{c d} \Psi_{a b} \Psi_{c d} = \\
& = & \sum_{a d} \Psi_{a d} \sum_{b} s(a,b,d) \xi_{a b} \zeta_{b d} := \sum_{a d} \rho_{a d} \Psi_{a d}
\end{eqnarray*}
having defined $\rho_{a d} = \sum_{b} s(a,b,d) \xi_{a b} \zeta_{b d}$.

This property shows also that EFB elements map directly to the isomorphic matrix algebra $\F( 2^m )$ where $a$ and $b$ are respectively the row and column indices of $\Psi_{a b}$ when interpreted as binary numbers substituting: $1 \to 0$ and $ -1 \to 1$. Let $e := (1,1,1,\ldots,1) \in \{ \pm 1 \}^m$ then, with the proposed substitutions, $e$ gives the binary expression of $0$ and $-e$ that of $2^m - 1$, see \cite{Budinich_2011_EFB}.

\section{Vector space $V$}
\label{Vector_space_V}
With the Witt basis (\ref{formula_Witt_basis}) it is easy to see that the null vectors $\{p_{i}\}$ can build vector subspaces made only of null vectors that we call Totally Null Planes (TNP, also: isotropic planes) of dimension at maximum $m$ \cite{Cartan_1937}. Moreover the vector space $V$ is easily seen to be the direct sum of two of these maximal TNP $P$ and $Q$ respectively:
\opt{margin_notes}{\mynote{mbh.note: NB given $P$ much freedom remains in choosing $q_i$'s}}
\begin{displaymath}
V = P \oplus Q \qquad
\left\{ \begin{array}{l l l}
P & := & \my_span{p_1, p_2, \ldots, p_m} \\
Q & := & \my_span{q_1, q_2, \ldots, q_m}
\end{array} \right.
\end{displaymath}
since $P \cap Q = \{0\}$ each vector $v \in V$ may be expressed in the form $v = \sum\limits_{i=1}^{m} \left( \alpha_{i} p_{i} + \beta_{i} q_{i} \right)$ with $\alpha_{i}, \beta_{i} \in \F$.
\opt{margin_notes}{\mynote{mbh.note: Very, very tricky point: see log. at pgs.: 443.6.1, 443.13.1, 492, 496 \& 502 ff.}}
Using (\ref{formula_Witt_basis_properties}) it is easy to derive the anticommutator of two generic vectors $v$ and $u = \sum\limits_{i=1}^{m} \left( \gamma_{i} p_{i} + \delta_{i} q_{i} \right)$
\begin{equation}
\label{v_u_commutator}
\anticomm{v}{u} = \sum\limits_{i=1}^{m} \alpha_{i} \delta_{i} + \beta_{i} \gamma_{i} \quad \in \F
\quad \Rightarrow \quad
\frac {1}{2} \anticomm{v}{v} = v^2 = \sum\limits_{i=1}^{m} \alpha_{i} \beta_{i} \dotinformula
\end{equation}
We define
$$
V_0 = \{ v \in V : v^2 = 0\} \qquad V_1 = \{ v \in V : v^2 \ne 0\}
$$
\opt{margin_notes}{\mynote{mbh.note: $0 \in V_0$ is coherent with $Q \subset V_0$ and $M(\Phi) = \{0\} \subset V_0$.}}
clearly $V = V_0 \cup V_1$ and $V_0 \cap V_1 = \emptyset$ but neither $V_0$ nor $V_1$ are subspaces of $V$ which is simple to see. Nevertheless $V_0$ contains subspaces of dimension $m$, \eg $Q$, and, similarly, $V_1$ contains subspaces of dimension $m$, \eg $\my_span{\gamma_1, \ldots, \gamma_{2 k - 1}, \ldots, \gamma_{2 m - 1}}$.
%Since a set cannot have dimension lower than those of one of its proper subsets we deduce $\dim_\F V_0 = \dim_\F V_1 = \dim_\F V/2 = m$.
%
\opt{margin_notes}{\mynote{mbh.note: more precisely one can only deduce $\dim_\F V_{0/1} \ge m$; commented offending phrase.}}

\begin{MS_Proposition}
\label{V0/1_prop}
Given any nonzero $v \in V$, there exists a nonzero spinor $\omega \in S$ such that $v \omega = 0$ if and only if $v \in V_0$. Conversely for any $v \in V_1$ and any nonzero $\omega \in S$ it follows $v \omega \ne 0$.
\end{MS_Proposition}
\begin{proof}
For any nonzero vector $v \in V_0$ we can take any $\omega \in S$, then either $v \omega = 0$ and $\omega$ is the spinor we search, or $v \omega \ne 0$, but then, since $S$ is a left ideal we have $\omega' := v \omega \in S$, it is not zero and $v \omega' = 0$. In turn for any $v$ such that $v \omega = 0$ it follows $v^2 \omega = 0$ but since $v^2 \in \F$ and $\omega \ne 0$ necessarily $v^2 = 0$. The second part is a direct consequence but we strengthen the result showing that given any $v \in V_1$ the existence of an hypothetical $\omega \in S$ such that $v \omega = 0$ leads to a contradiction. Let's suppose such $\omega$ exists, from $v \omega = 0$ we get $v^2 \omega = 0$ and, since $v^2 \ne 0$, this would imply $\omega = 0$.
\end{proof}

\subsection{Conjugation in $V$}
\label{Conjugationi_V}
When $\F = \C$ complex conjugation in vector space $V$ is given by
\begin{equation}
\label{v_vbar}
v = \sum_{i = 1}^m \alpha_i p_i + \beta_i q_i \quad \Rightarrow \quad \myconjugate{v} = \sum_{i = 1}^m \myconjugate{\beta}_i p_i + \myconjugate{\alpha}_i q_i
\end{equation}
that with (\ref{v_u_commutator}) gives $\myconjugate{v}^2 = \myconjugate{v^2}$. For $\F = \R$, since $\myconjugate{\alpha}_i = \alpha_i$, the conjugation is obtained by exchanging basis vectors $p_i$ and $q_i$ (or, identically, exchanging coefficients $\alpha_i$ and $\beta_i$) and in both cases conjugation defines an involutive automorphism on $V$ since $\myconjugate{\myconjugate{v}} = v$;

For $\F = \R$ we can go further: by (\ref{v_u_commutator}) $\myconjugate{v}^2 = v^2$ and this conjugation is an isometry on $V$ that lifts uniquely to an automorphism on the entire algebra and since our algebra is central simple all its automorphisms are inner. So there must exist $C$ such that $\myconjugate{v} = C v C^{-1}$.

To find its explicit form let $\Delta_\pm = (p_1 \pm q_1) \cdots (p_m \pm q_m)$ and with (\ref{formula_Witt_basis}) it is easy to see that $\Delta_+ = \gamma_1 \cdots \gamma_{2 k - 1} \cdots \gamma_{2 m - 1}$ whereas $\Delta_-$ is the product of the even, spacelike, $\gamma$'s. With (\ref{space_signature}) one easily finds $\Delta_\pm^2 = (-1)^{\frac{m (m \mp 1)}{2}}$ and defining
\begin{equation}
\label{C_def}
C =
\left\{ \begin{array}{l}
\Delta_+\\
\Delta_-
\end{array} \right.
\qquad C^{-1} =
\left\{ \begin{array}{l}
(-1)^{\frac{m (m - 1)}{2}} \Delta_+ \qquad \mbox{for} \; m \; \mbox{odd}\\
(-1)^{\frac{m (m + 1)}{2}} \Delta_- \qquad \mbox{for} \; m \; \mbox{even}
\end{array} \right.
\end{equation}
we can prove that $\myconjugate{v} = C v C^{-1}$: it suffices to write $v$ in the Witt basis and make the simple exercise of proving that $C p_i C^{-1} = q_i$.
\opt{margin_notes}{\mynote{mbh.ref: see p. 273 and on Spinorial Chessboard p. 106. The detailed proof follows and is commented.}}
%
%
%Let's suppose that $m$ is odd, then $\myconjugate{v} = (-1)^{\frac{m (m - 1)}{2}} \Delta_+ v \Delta_+$ with which
%$$
%(-1)^{\frac{m (m - 1)}{2}} \Delta_+ v \Delta_+ = (-1)^{\frac{m (m - 1)}{2}} \sum_{i = 1}^m \alpha_i \Delta_+ p_i \Delta_+ + \beta_i \Delta_+ q_i \Delta_+
%$$
%and, since $(p_i + q_i) p_i (p_i + q_i) = q_i$,
%\begin{eqnarray*}
%(-1)^{\frac{m (m - 1)}{2}} \Delta_+ p_i \Delta_+ & = & (-1)^{\frac{m (m - 1)}{2}} (p_1 + q_1) \cdots (p_m + q_m) p_i (p_1 + q_1) \cdots (p_m + q_m) = \\
%& = & (p_m + q_m) \cdots (p_1 + q_1) p_i (p_1 + q_1) \cdots (p_m + q_m) = \\
%& = & (-1)^{i-1} (p_m + q_m) \cdots (p_i + q_i) p_i (p_i + q_i) \cdots (p_m + q_m) = \\
%& = & (-1)^{m-1} q_i = q_i
%\end{eqnarray*}
%since, by hypothesis, $m$ is odd. Similarly $(-1)^{\frac{m (m - 1)}{2}} \Delta_+ q_i \Delta_+ = p_i$ so that the proposition is proved; for $m$ even the demonstration is very similar.
%
One easily verifies
$$
\myconjugate{\myconjugate{v}} = C C v C^{-1} C^{-1} = C C^{-1} v C C^{-1} = v \dotinformula
$$

Returning to the case $\F = \C$, it is obvious that also in this case $C$ can be defined and $C p_i C^{-1} = q_i$ so that, indicating with $v^\star$ the vector $v$ with complex conjugate field coefficients, we can write (\ref{v_vbar}) as
\begin{equation*}
\label{v_vbar_generalized}
\myconjugate{v} = C v^\star C^{-1}
\end{equation*}
that holds also for $\F = \R$ since in this case $v^\star = v$ and thus from now on we will stick to this form for (complex) conjugation. It is an easy exercise to verify that this form generalizes to any element of the algebra $\omega$ giving
\opt{margin_notes}{\mynote{mbh.ref: for precise proofs see log. p. 502.3 and 507}}
$$
\myconjugate{\omega} = C \omega^\star C^{-1}
$$
and that, for both $\F = \C$ and $\R$,
\begin{equation*}
\label{v0_equal_v0bar}
v^2 = 0 \iff \myconjugate{v}^2 = 0 \dotinformula
\end{equation*}

\begin{MS_Proposition}
\label{v_vbar_prop}
Given nonzero vector $v$ and $\omega \in S$ such that $v \omega = 0$ it follows $\myconjugate{v} \omega \ne 0$, conversely $\myconjugate{v} \omega = 0$ implies $v \omega \ne 0$.
\end{MS_Proposition}
\begin{proof}
We start showing that for any nonzero vector $v$ and for both $\F = \R$ and $\C$ one has $(v + \myconjugate{v})^2 > 0$. With (\ref{v_vbar}) one easily finds that $v + \myconjugate{v} = \sum_{i = 1}^m \gamma_i p_i + \myconjugate{\gamma}_i q_i$ and with (\ref{v_u_commutator}) $(v + \myconjugate{v})^2 = \sum_{i = 1}^m \gamma_i \myconjugate{\gamma}_i > 0\,$%
\footnote{note that also $(v - \myconjugate{v})^2 < 0$}%
. With proposition~\ref{V0/1_prop} it follows that for any vector $v$: $(v + \myconjugate{v}) \omega \ne 0$ that, if one of the terms is zero, implies that the other must be nonzero.
\end{proof}
We remark that this result is just an implication holding only when one of the two terms $v \omega$ or $\myconjugate{v} \omega$ is zero since there are cases in which both terms can be nonzero, \eg $v = p_1$, $\omega = q_1 q_2 \cdots q_m + p_1 q_1 q_2 \cdots q_m$.
\opt{margin_notes}{\mynote{mbh.ref: expanded at log-book pg. 441.}}

\section{Spinor spaces}
\label{Spinor_spaces}
We have seen in section~\ref{EFB_formalism} that \myCl{m}{m}{g}, as a vector space, is the direct sum of subspaces of different $h \circ g-$signatures. Given the Clifford product properties (\ref{EFB_products}) these subspaces are also minimal left ideals of \myCl{m}{m}{g} and thus coincide with $2^m$ different spinor spaces $S_{h \circ g}$ (that in turn correspond to different columns of the isomorphic matrix algebra $\F( 2^m )$). We choose the spinor space with $h \circ g = - e$ so that when we speak of a generic $S$ we refer to the particular spinor space $S_{-e}$ used to build the Fock basis \cite{BudinichP_1989}. Its generic element is described by: $\omega = \sum_{a} \xi_{a b} \Psi_{a b}$ and, since the second index of the $h \circ g-$signature is constant, whenever possible we will omit it, writing for the spinor expansion in the Fock basis
\begin{equation}
\label{Fock_basis_expansion}
\omega \in S \qquad \omega = \sum_{a} \xi_{a} \Psi_{a} \dotinformula
\end{equation}

Here we are interested mainly in the relations between spinors and TNP and we try to investigate them independently of the particular basis.

For each nonzero spinor $\omega \in S$ we define its associated TNP as:
$$
M(\omega) := \{v \in V : v \omega = 0
% \mbox{ and } \anticomm{v_i}{v_j} = 0 \quad \forall v_i, v_j \in M(\omega)
\}
$$
and the spinor is \emph{simple} iff the TNP is of maximal dimension, \ie iff $\dim_\F M(\omega) = m$. It is easy to see that all vectors in $M(\omega)$ are mutually orthogonal and that $M(\omega)$ is a vector subspace of $V$ contained in $V_0$.

Since all EFB elements are simple spinors each of them has an associated TNP of maximal dimension uniquely identified by the $h-$signature $a$ of $\Psi_{a}$; for example if $a = (-1,1,1, \ldots, 1)$ then $\Psi_a := \Psi_{(-1,1,1, \ldots, 1)} = p_1 q_1 q_2 q_3 \cdots q_m$ and $M(\Psi_a) = \my_span{ p_1, q_2, q_3, \ldots, q_m}$.

\begin{MS_Proposition}
\label{omega_omegabar_prop}
For any nonzero vector $v$ and $\omega \in S$ such that $v \omega = 0$ it follows $v \myconjugate{\omega} \ne 0$, conversely $v \myconjugate{\omega} = 0$ implies $v \omega \ne 0$.
\end{MS_Proposition}
\begin{proof}
By proposition~\ref{v_vbar_prop} we know that $v \omega = 0$ implies $\myconjugate{v} \omega \ne 0$ and thus
$$
0 \ne \myconjugate{\myconjugate{v} \omega} = \myconjugate{\myconjugate{v}} {\,} \myconjugate{\omega} = v \myconjugate{\omega} \dotinformula
%\myconjugate{v} \omega = C v C^{-1} \omega = C^{-1} v C \omega C^{-1} C = C^{-1} v \myconjugate{\omega} C \ne 0
$$
Similarly from $v \myconjugate{\omega} = 0$ by propositions~\ref{v_vbar_prop} one obtains $\myconjugate{v} {\,} \myconjugate{\omega} \ne 0$ and thus $v \omega \ne 0$.
\end{proof}
\begin{MS_Corollary}
\label{omega_omegabar_coro}
For any nonzero $v \in V_0$, given nonzero $\omega \in S$ such that $v \omega = 0$ it follows $v C \omega^\star \ne 0$, conversely $v C \omega^\star = 0$ implies $v \omega \ne 0$.
\end{MS_Corollary}
\begin{proof}
By proposition~\ref{v_vbar_prop} we know that $v \omega = 0$ implies $\myconjugate{v} \omega = C^{-1} v^\star C \omega \ne 0$ and since $S$ is a minimal left ideal it follows $\myconjugate{v} \omega \in S$. Then since $C$ is made of vectors with length $\pm 1$ and with proposition~\ref{V0/1_prop}, we get $0 \ne C \myconjugate{v} \omega = v^\star C \omega$ and also the ``starred'' form of this relation is nonzero thus $0 \ne (v^\star C \omega)^\star = v C \omega^\star$. The other case is similar.
\opt{margin_notes}{\mynote{mbh.note: for calculations of $C \omega$ see log p. 394.32'; for $s(a)$ see log p. 460, for starred relations see p. 502.3}}
\end{proof}

\subsection{The ``generic'' spinor $\Phi$}
\label{Generic_spinor}
Given the spinor expansion (\ref{Fock_basis_expansion}) we call $\Phi$ the ``generic'' spinor of $S$
\begin{equation}
\label{Phi_def}
\Phi := \sum_a \xi_a \Psi_a
\end{equation}
with the understanding that the field coefficients $\xi_a$ are taken as ``indeterminates'' \ie that they are free to take any value; varying the coefficients $\Phi$ spans the entire $S$ so when writing $\Phi$ we will substantially refer to $S$. $\Phi$ will be said to be in \emph{general position} when all field coefficients $\xi_a$ are nonzero \cite{Trautman_Trautman_1994}.

This variability of the coefficients is a critical point: as a rule of thumb one can say that varying the values of the field coefficients does not alter the properties of a spinor \emph{as long as} they remain different from zero. We explain this with two examples: let $\omega := v \Phi \ne 0$ where $v \in V_0$; obviously $v \omega = v^2 \Phi = 0$ and this happens for any choice of the coefficients $\xi_a$ in $\Phi$ showing that, at least as far as these properties of the spinor are concerned, the particular values of the coefficients are irrelevant. To show that $0$ is a critical value we consider another example in \myCl{2}{2}{g}: let us take $\omega = \xi_1 p_1 q_1 q_2 + \xi_3 p_1 q_1 p_2 q_2$; it is simple to see that $v = \alpha p_1$ for any $\alpha \in \F$ are the only vectors such that $v \omega = 0$ and this is true for any value of the coefficients $\xi_1, \xi_3$. But if $\xi_1 = 0$ then another null vector annihilates $\omega$ since $p_1 \omega = p_2 \omega = 0$, similarly if $\xi_3 = 0$ then $p_1 \omega = q_2 \omega = 0$. These examples show that we are moving along a treacherous path and that one must proceed with some care. For a spinor in general position with $m \ne 2$, $v \Phi = 0$ only iff $v = 0$ \cite{Trautman_Trautman_1994, Budinich_2014} so we can assume
\opt{margin_notes}{\mynote{mbh.note: this does not hold for $m = 2$, see log. p. 489}}
$$
M(\Phi) = \{0\} \qquad {\rm and} \qquad \dim_\F M(\Phi) = 0
$$
and this enriches the correspondences between $V_0$ and $S$: any null vector $v$ identifies the annihilating spinors (see an explicit construction in the proof of proposition~\ref{V0/1_prop}). Conversely, almost any spinor annihilates one or more null vectors, an exception is $\Phi$ but it is not the only one.

\begin{MS_Proposition}
\label{S_bisect_prop}
\opt{margin_notes}{\mynote{mbh.ref: for a different proof see log. p. 513}}
Any nonzero $v \in V_0$ partitions the spinor space $S$ into two subsets: $S_v = \{\omega \in S : v \omega = 0\}$ and $\myconjugate{S}_{v} = \{\omega \in S : v \omega \ne 0\}$ so that for any $v$, $S_{v} \cap \myconjugate{S}_{v} = \emptyset$ and $S_{v} \cup \myconjugate{S}_{v} = S$. Moreover let $S_{\myconjugate{v}} = \{\omega \in S : \myconjugate{v} \omega = 0\}$, the following hold:
\begin{itemize}
\item $S_v$ and $S_{\myconjugate{v}}$ are subspaces of $S$ and $S_v \cap S_{\myconjugate{v}} = \{0\}$,
\item $S_{\myconjugate{v}} \subset \myconjugate{S}_{v}$,
\item $\dim_\F S_{v} = \dim_\F S_{\myconjugate{v}} = 2^{m - 1}$,
\item $S = S_{v} \oplus S_{\myconjugate{v}}$ .
\end{itemize}
\end{MS_Proposition}
\begin{proof}
We start showing that $S_{v}$ and $\myconjugate{S}_{v}$ are both non empty: given any nonzero $v \in V_0$ and $\omega \in S$, $v \omega$ is either zero or not. If $v \omega = 0$ then $\omega \in S_{v}$ and, by corollary~\ref{omega_omegabar_coro}, $\omega' := C \omega^\star \in \myconjugate{S}_{v}$; if $v \omega \ne 0$ then $\omega \in \myconjugate{S}_{v}$ and $\omega' := v \omega \in S_{v}$. It is also obvious that the $S_{v}$ and $\myconjugate{S}_{v}$ partition $S$ since any $\omega \in S$ it is either in $S_{v}$ or in $\myconjugate{S}_{v}$.

For any $\omega \in S_{\myconjugate{v}}$ we get by proposition~\ref{v_vbar_prop} $v \omega \ne 0$ and thus $S_{\myconjugate{v}} \subset \myconjugate{S}_{v}$ moreover the inclusion is strict since there exists spinors such that both $v \omega \ne 0$ and $\myconjugate{v} \omega \ne 0$ as shown in the example after proposition~\ref{v_vbar_prop}. It is also simple to see that both $S_v$ and $S_{\myconjugate{v}}$ are vector subspaces of $S$ and that, by proposition~\ref{v_vbar_prop}, $S_v \cap S_{\myconjugate{v}} = \{0\}$.

To prove the statement about dimension we start proving that for any $\omega$ in one subspace there exists a ``twin'' spinor $\omega'$, linearly independent from $\omega$, belonging to the other subspace. Let's suppose first $\omega \in S_{v}$ then, by proposition~\ref{v_vbar_prop}, $\omega' := \myconjugate{v} \omega \ne 0$ and since $\myconjugate{v} \omega' = \myconjugate{v}^2 \omega = 0$ then $\omega' \in S_{\myconjugate{v}}$. Moreover $\omega'$ is linearly independent from $\omega$ since the hypothesis $\omega = \alpha \omega'$ is in contradiction with $v \omega = 0$, $v \omega' \ne 0$. If the initial spinor $\omega$ is in $S_{\myconjugate{v}}$ then $\omega' := v \omega \ne 0$ and $v \omega' = v^2 \omega = 0$ and thus $\omega' \in S_{v}$ and is linearly independent from $\omega$. Every spinor lying in one subspace has thus a linearly independent twin in the other subspace that implies $\dim_\F S_{v} = \dim_\F S_{\myconjugate{v}}$.

We prove now that $S_{v} \oplus S_{\myconjugate{v}} = S$ and thus $\dim_\F S_{v} = \dim_\F S_{\myconjugate{v}} = 2^{m - 1}$. To do this we perform a proper rotation in vector space $V$ such that the null vectors $v$ and $\myconjugate{v}$ are transformed, respectively, to $q_1$ and $p_1$ of the new Witt basis of $V$. Building the associated EFB of $S$ we get that in the expansion (\ref{Phi_def}) any spinor $\omega$ can have only components with $h_1 = \pm1$ that correspond to spinors of $S_{v}$ or of $S_{\myconjugate{v}}$ and thus $S_{v} \oplus S_{\myconjugate{v}} = S$.
\end{proof}

We remark that while $S_{v}$ is a vector subspace of $S$, $\myconjugate{S}_{v}$ is not a subspace: consider again an example in \myCl{2}{2}{g}: $v = p_1 + q_2$ and $\Psi_0 = q_1 q_2$, $\Psi_3 = p_1 q_1 p_2 q_2$. Clearly $v^2 = 0$ and $v \Psi_0 = v \Psi_3 = p_1 q_1 q_2$ but $v (\Psi_0 - \Psi_3) = 0$.

We now introduce the notation $v \Phi$ where $v$ is a nonzero vector of $V_0$ and $\Phi$ is the generic spinor (\ref{Phi_def}). Consider for example $v = q_i$, when we calculate $v \Phi$ all the terms of the expansion (\ref{Phi_def}) in which $h_i = 1$ (\ie those $\Psi_a = \cdots q_i \cdots$) are immediately set to $0$ independently of the values of the coefficients $\xi_a$. So with $q_i \Phi$ we indicate the generic spinor with $h_i = 1$, \ie a spinor with only half of the elements of the Fock basis. So in general with $v \Phi$ we mean the generic spinor whose components have ``survived'' to the multiplication by $v$. In the following proposition we show that this property of halving the spinor space spanned by $\Phi$ does not depend on the particular choice $v = q_i$ but is general.

\begin{MS_Proposition}
\label{generic_multiple_v_spinor}
Given $k \le m$ nonzero $v_1, v_2, \ldots, v_{k} \in V_0$ forming a TNP of dimension $k$, any spinor that annihilates $v_1, v_2, \ldots, v_{k}$ can be written $v_1 v_2 \cdots v_{k} \Phi$, \ie one can write $S_{v_1, v_2, \ldots, v_{k}} = v_1 v_2 \cdots v_{k} \Phi$ and $\dim_\F S_{v_1, v_2, \ldots, v_{k}} = 2^{m - k}$.
\opt{margin_notes}{\mynote{mbh.note: a similar prop appeared in M\&P Graph\_514 prop 2 (15) p. 8 but this is much better. This proposition appears not to be inficiated by results at log p. $503.2$ ff. showing that there are no spinors with $N(\omega) = m - 2$ (we are \emph{not} saying that $N(v_1 v_2 \cdots v_{k} \Phi) = k$ but just that $N(v_1 v_2 \cdots v_{k} \Phi) \ge k$) see p. 511.}}
\end{MS_Proposition}
\begin{proof}
The proof is by induction on $k$ so we first prove the case with $k = 1$: we start showing that for any $\omega \in S_{v}$ there exists $\omega^{''} \in S_{\myconjugate{v}}$ such that $\omega = v \omega^{''}$. In previous proof we saw that $\omega' := \myconjugate{v} \omega \ne 0$ is such that $v \omega' \ne 0$ and
$$
v \omega' = v \myconjugate{v} \omega = \anticomm{v}{\myconjugate{v}} \omega = \alpha \omega
$$
where $\alpha = \anticomm{v}{\myconjugate{v}} \in \F - \{0\}$ by hypothesis. So defining $\omega^{''} = \alpha^{-1} \omega' = \alpha^{-1} \myconjugate{v} \omega$ we get $v \omega^{''} = \omega$ and clearly $\omega^{''} \in S_{\myconjugate{v}} \subset S$. If we set the coefficients $\xi_a$ of (\ref{Phi_def}) to get $\Phi = \omega^{''}$ we will obtain $v \Phi = \omega$. Since this procedure works for any $\omega \in S_{v}$ we have thus proved that $v \Phi$ can reach any $\omega \in S_{v}$ and thus that $S_{v} \subseteq v \Phi$. On the other hand for any $\omega \in v \Phi$ one has $v \omega = v^2 \Phi = 0$ and thus $S_{v} = v \Phi$.

This means that the most general spinor that annihilates $v \in V_0$ can always be written, for an appropriate choice of the coefficients $\xi_a$, as $v \Phi$. With proposition~\ref{S_bisect_prop} follows immediately: $\dim_\F v \Phi = \dim_\F S_{v} = 2^{m - 1}$ that generalizes the result, mentioned before, that $q_i \Phi$, spans a $2^{m - 1}$-dimensional space.

For the induction step we suppose that any spinor annihilating $v_1, v_2, \ldots, v_{k-1}$ may be written, with an appropriate choice of the coefficients $\xi_a$ in (\ref{Phi_def}), as $v_1 v_2 \cdots v_{k-1} \Phi$ and thus $S_{v_1, v_2, \ldots, v_{k-1}} = v_1 v_2 \cdots v_{k-1} \Phi$ and $\dim_\F S_{v_1, v_2, \ldots, v_{k-1}} = 2^{m - k + 1}$.

Let us suppose that we add a new $k$-th vector and that our $k$ vectors $v_1, v_2, \ldots, v_{k}$ form a basis of the TNP obeying the standard relations (\ref{formula_Witt_basis_properties}):
$$
\anticomm{v_{i}}{v_{j}} = \anticomm{\myconjugate{v}_{i}}{\myconjugate{v}_{j}} = 0
\qquad
\anticomm{v_{i}}{\myconjugate{v}_{j}} = \delta_{i j} \qquad 1 \le i,j \le k
$$
that can always be obtained by a proper rotation in $\my_span{v_1, v_2, \ldots, v_{k}}$ since the vectors are linearly independent by hypothesis (we will show in the next proposition that this hypothesis is not a limitation).

Let's now take any $\omega \in S_{v_1, v_2, \ldots, v_{k}}$, clearly $v_{k} \omega = 0$ but, by proposition~\ref{v_vbar_prop}, $\omega' := \myconjugate{v}_{k} \omega \ne 0$, from which $\myconjugate{v}_{k} \omega' = 0$ from which $v_{k} \omega' \ne 0$. But, since $\anticomm{v_i}{\myconjugate{v}_{k}} = 0$ for $i = 1, \ldots, k-1$ it follows that $v_i \omega' = v_i \myconjugate{v}_{k} \omega = - \myconjugate{v}_{k} v_i \omega = 0$ for $i = 1, \ldots, k-1$ and thus $\omega' \in S_{v_1, v_2, \ldots, v_{k-1}}$ and thus, by induction hypothesis, for appropriate coefficients $\xi_a$, we have $\omega' = v_1 v_2 \cdots v_{k-1} \Phi$.

We know $v_{k} \omega' \ne 0$ and that $\anticomm{v_{k}}{\myconjugate{v}_{k}} = 1$ thus
$$
v_{k} \omega' = v_{k} \myconjugate{v}_{k} \omega = \anticomm{v_{k}}{\myconjugate{v}_{k}} \omega = \omega
$$
and since $\omega'$ is already written in the form $v_1 v_2 \cdots v_{k-1} \Phi$ we derive that also any $\omega \in S_{v_1, v_2, \ldots, v_{k}}$ may be written as $\omega = v_{k} \omega' = v_{k} v_1 v_2 \cdots v_{k-1} \Phi = (-1)^{k - 1} v_1 v_2 \cdots v_{k-1} v_{k} \Phi$ since $\anticomm{v_i}{v_k} = 0$ for any $1 \le i \le k-1$. Thus $S_{v_1, v_2, \ldots, v_{k}} \subseteq v_1 v_2 \cdots v_{k} \Phi$ and since any $\omega \in v_1 v_2 \cdots v_{k} \Phi$ is necessarily also in $S_{v_1, v_2, \ldots, v_{k}}$ it follows $S_{v_1, v_2, \ldots, v_{k}} = v_1 v_2 \cdots v_{k} \Phi$.
\opt{margin_notes}{\mynote{mbh.note: tricky point}}

To prove the statement about dimension one can use the previous argument of the twin spinors to show that in $S_{v_1, v_2, \ldots, v_{k-1}}$ there are two subspaces of spinors of equal dimension: one annihilates $v_{k}$ and the other annihilates $\myconjugate{v}_{k}$ and since their sum has dimension $2^{m - k + 1}$ it follows that the first subspace, \ie $S_{v_1, v_2, \ldots, v_{k}}$, has dimension $2^{m - k}$.
\end{proof}

An immediate consequence of this result is that any simple spinor $\omega \in S$ may be written as $\omega = v_1 v_2 \cdots v_{m} \Phi$ where $\my_span{v_1, v_2, \ldots, v_{m}} = M(\omega)$, $S_{v_1, v_2, \ldots, v_{m}} = v_1 v_2 \cdots v_{m} \Phi$ and $\dim_\F S_{v_1, v_2, \ldots, v_{m}} = 1$, \ie all simple spinors form one-dimensional subspaces of $S$. it is simple to see that the converse is not true in general; moreover in \cite{Budinich_2011_EFB} it is shown that in any basis a simple spinor can have at most $m$ nonzero coordinates in (\ref{Phi_def}).
\opt{margin_notes}{\mynote{mbh.note: the converse is not true: there are one-dimensional subspaces of $S$ that are not simple spinors \eg the example (commented) at the end of section~\ref{B_product}.}}

We show now that the choice of the null vectors $v_1, v_2, \ldots, v_{k}$ used to define $\omega := v_1 v_2 \cdots v_{k} \Phi$ is completely free provided they define the very same $M(\omega)$.

\begin{MS_Proposition}
\label{free_v1_vk}
The generic spinor $\omega := v_1 v_2 \cdots v_{k} \Phi$ with $M(\omega) = \my_span{v_1, v_2, \ldots, v_{k}}$, changes only by a multiplicative constant if the defining vectors are changed to $v_1', v_2', \ldots, v_{k}'$ spanning the same $M(\omega)$. The multiplicative constant is the determinant of the matrix $A$ transforming $v_i$ to $v_i'$.
\opt{margin_notes}{\mynote{mbh.note: see log-book pg. 356.13., same result also in Math\_532 p. 107}}
\end{MS_Proposition}
\begin{proof}
Given a proper linear transformation $A$ changing $v_i$ to $v_i'$ it is easy to see that
$$
\omega' := v_1' v_2' \cdots v_{k}' \Phi = \left(\sum_{i = 1}^k a_{1 i} v_i \right) \left(\sum_{i = 1}^k a_{2 i} v_i \right) \cdots \left(\sum_{i = 1}^k a_{k i} v_i \right) \Phi
$$
and expanding the product of sums it is clear that all the terms involving powers greater than $1$ of any $v_i$ are zero since all the vectors $v_i$ are null. It follows that of the initial $k^k$ terms in $\omega'$ only the $k!$ terms of the form $v_{\pi_1} v_{\pi_2} \cdots v_{\pi_k}$, where $({\pi_1}, {\pi_2}, \ldots, {\pi_k})$ is a permutation of $(1,2, \ldots, k)$ survive. Given that $v_i v_j = -v_j v_i$ for any $i \ne j$ it follows that all the terms can be brought to the form $ \pm v_1 v_2 \cdots v_{k}$. We conclude showing that actually
$$
\omega' = v_1' v_2' \cdots v_{k}' \Phi = \det{A} \, v_1 v_2 \cdots v_{k} \Phi = \det{A} \, \omega \dotinformula
$$
We proceed by induction: for $k = 2$ we have
\opt{margin_notes}{\mynote{mbh.note: see log-book pg. 263 \& 356.13.}}
$$
\omega' = v_1' v_2' \Phi = (a_{11} v_1 + a_{12} v_2) (a_{21} v_1 + a_{22} v_2) \Phi = (a_{11} a_{22} - a_{12} a_{21}) v_1 v_2 \Phi = \det{A} \, \omega
$$
the induction step follows easily from simple determinant properties.
\end{proof}

With these last two propositions we can generalize the concept of generic spinor (\ref{Phi_def}) from $\Phi$, the generic spinor with $M(\Phi) = \{0\}$, to $\omega := v_1 v_2 \cdots v_{k} \Phi$ that is the generic spinor having $M(\omega) = \my_span{ v_1, v_2, \ldots, v_{k}}$; moreover the choice of the null vectors $v_1, v_2, \ldots, v_{k}$ used to define $\omega := v_1 v_2 \cdots v_{k} \Phi$ is completely free.

\subsection{The inner product $\langle B \cdot , \cdot \rangle$ of spinor spaces}
\label{B_product}
We now use these results to give different proofs of some known results and to prove some new ones but we start with a concise summary.
\opt{margin_notes}{\mynote{mbh.probs: to be refined... define properly $v$, $v^{t}$ and $\myconjugate{v}$, see also p. 6. See at page 386.}}

The transposed generators (endomorphisms) $\gamma_i^{t}$ admit a representation of \myCl{m}{m}{g} in $S^*$, the dual of $S$. Since \myCl{m}{m}{g} is simple, there is an isomorphism $B: S \to S^*$ interwining the representations (see \cite{BudinichP_1989} and \cite{Case_1955})
\begin{equation}
\label{B_op_def}
\gamma_i^{t} B = B \gamma_i \qquad \mbox{and} \qquad B^{t} = (-1)^{\frac{m (m-1)}{2}} B \dotinformula
\end{equation}
The isomorphism $B$ defines also an inner product ($\langle \cdot , \cdot \rangle$ represents the bilinear product)
$$
S \times S \to \F \qquad B(\omega, \varphi) := \langle B \omega, \varphi \rangle \in \F
$$
which is invariant with respect to the action of the group \Pin{g} made of unit vectors \ie vectors $v$ such that $v^2 = 1$, namely:
$$
B(v \omega, v \varphi) = \langle B v \omega, v \varphi \rangle = \langle v^{t} B \omega, v \varphi \rangle = \langle B \omega, v^2 \varphi \rangle = B(\omega, \varphi) \dotinformula
$$

We now generalize proposition III.2.4 of \cite{Chevalley_1954} relaxing partially the demanding condition of spinors being simple, while, at the same time, giving a simpler proof:
\begin{MS_Proposition}
\label{B_omega_varphi}
For any nonzero spinors $\omega, \varphi \in S$ with $\dim_\F M(\omega) > 0$ and $\dim_\F M(\varphi) > 0$ then $M(\omega) \cap M(\varphi) \ne \{0\}$ implies $B(\omega, \varphi) = 0$.

Viceversa given nonzero spinors $\omega, \varphi \in S$ with $\dim_\F M(\omega) = m$ and $\dim_\F M(\varphi) > m - 3$ then $B(\omega, \varphi) = 0$ implies $M(\omega) \cap M(\varphi) \ne \{0\}$.
\opt{margin_notes}{\mynote{mbh.note: this is the \emph{correct} version of the proposition, see a discussion at logbook pp. 526 ff.}}
\end{MS_Proposition}
\begin{proof}
Let's suppose first that $v \in M(\omega) \cap M(\varphi)$, then $v \omega = v \varphi = 0$. Let's ``normalize'' $v$ such that $\anticomm{v}{\myconjugate{v}} = 1$, then,
$$
\langle B \omega, \varphi \rangle = \langle B \omega, \anticomm{v}{\myconjugate{v}} \varphi \rangle = \langle B \omega, v \myconjugate{v} \varphi \rangle = \langle v^{t} B \omega, \myconjugate{v} \varphi \rangle = \langle B v \omega, \myconjugate{v} \varphi \rangle = 0 \dotinformula
$$

To prove the second part let's suppose $B(\omega, \varphi) = 0$ and $M(\omega) = \my_span{v_1, v_2, \ldots, v_{m}}$, $M(\varphi) = \my_span{u_1, u_2, \ldots, u_{l}}$ with $l > m - 3$. We start observing that assuming in full generality that $\omega = \Psi_{e}$ and for any $\varphi \in S$ expanded with (\ref{Fock_basis_expansion}) we have
\begin{equation}
\label{in_this_proof}
0 = B(\omega, \varphi) = \sum_{a} \xi_{a} B(\Psi_{e}, \Psi_{a}) = \xi_{-e} B(\Psi_{e}, \Psi_{-e})
\end{equation}
where the last equality derives from the forward part of this proposition, so that we can conclude that necessarily $\xi_{-e} = 0$. This is enough to prove the thesis when the 2 spinors are simple and $l = m$ since taking \eg $\varphi = \Psi_{a}$ (that is always possible, see proposition~2 of \cite{BudinichP_1989}) any $\Psi_{a} \ne \Psi_{-e}$ has $M(\Psi_{e}) \cap M(\Psi_{a}) \ne \{0\}$. Let's suppose now $l = m - 1$ and with proposition~\ref{generic_multiple_v_spinor} we may write
$$
\varphi = u_1 u_2 \cdots u_{m - 1} \Phi = u_1 u_2 \cdots u_{m - 1} (\xi_1 u_m + \xi_2 \myconjugate{u}_m u_m)
$$
where the last equality can be easily explained assuming that $\my_span{u_1, u_2, \ldots, u_{m}}$ form a MTNP and that the very same vectors form a Fock basis of $S$. Since
$$
0 = B(\omega, \varphi) = B\left(v_1 v_2 \cdots v_{m} \Phi, u_1 u_2 \cdots u_{m - 1} (\xi_1 u_m + \xi_2 \myconjugate{u}_m u_m) \right)
$$
if $\my_span{v_1, v_2, \ldots, v_{m}} \cap \my_span{u_1, u_2, \ldots, u_{m - 1}} \ne \{ 0 \}$ the proposition is satisfied. It remains the case $\my_span{v_1, v_2, \ldots, v_{m}} \cap \my_span{u_1, u_2, \ldots, u_{m - 1}} = \{ 0 \}$ that implies $\my_span{u_1, u_2, \ldots, u_{m - 1}} \subset \my_span{\myconjugate{v}_1, \myconjugate{v}_2, \ldots, \myconjugate{v}_{m}}$.

Supposing \eg that also $u_m \in \my_span{\myconjugate{v}_1, \myconjugate{v}_2, \ldots, \myconjugate{v}_{m}}$, by (\ref{in_this_proof}) it follows that in this case $B(\omega, \varphi) = 0$ requires $\xi_1 = 0$ that, in turn, since $\varphi \ne 0$, implies $\xi_2 \ne 0$. So in this case the hypothesis $B(\omega, \varphi) = 0$ implies that $\varphi = \xi_2 u_1 u_2 \cdots u_{m - 1} \myconjugate{u}_m u_m$ and thus $\myconjugate{u}_m \in M(\omega) \cap M(\varphi)$ proving the proposition for $l = m - 1$.

The proof of the case $l = m - 2$ is very similar, we start by writing
\begin{eqnarray*}
\varphi & = & u_1 u_2 \cdots u_{m - 2} \Phi = \\
& = & u_1 u_2 \cdots u_{m - 2} (\xi_1 u_{m - 1} u_m + \xi_2 u_{m - 1} \myconjugate{u}_m u_m + \xi_3 \myconjugate{u}_{m - 1} u_{m - 1} u_m + \xi_4 \myconjugate{u}_{m - 1} u_{m - 1} \myconjugate{u}_m u_m)
\end{eqnarray*}
and as before $B(\omega, \varphi) = 0$ implies \eg $\xi_1 = 0$ and it's an easy exercise to show that for any choice of $\xi_2, \xi_3, \xi_4$ then $v' = \xi_3 \myconjugate{u}_{m - 1} - \xi_2 \myconjugate{u}_{m}$ is null and belongs to $M(\omega) \cap M(\varphi)$.
\end{proof}
We remark that the proposition is strict in the sense that is easy to find counterexamples with $\dim_\F M(\omega) = m$, $\dim_\F M(\varphi) = m - 3$ or $\dim_\F M(\omega) = \dim_\F M(\varphi) = m - 1$, and $B(\omega, \varphi) = 0$ with $M(\omega) \cap M(\varphi) = \{0\}$.

\section{Simple spinors}
\label{Simple_spinors}

We start remembering that the endomorphisms of $S$, $\End_{\F} S$, provide the representations of \myCl{m}{m}{g} and with the canonical isomorphism $\End_{\F} S \cong S \otimes S^*$ any $\mu \in \myCl{m}{m}{g}$ can be written as $\mu \cong \omega \otimes \varphi^*$ for $\omega, \varphi \in S$ and its action on any spinor $\phi \in S$ is given by
\opt{margin_notes}{\mynote{mbh.note: done at pages 387, 389'.}}
$$
\mu (\phi) = \omega \otimes \varphi^* (\phi) := \langle \varphi^*, \phi \rangle \omega = \langle B \varphi, \phi \rangle \omega
$$
and since any $\mu \in \myCl{m}{m}{g}$ can also be expressed in a standard multivector expansion
\begin{equation}
\label{multivector_expansion}
\mu \cong \omega \otimes \varphi^* = \sum_{k = 0}^{2 m} \sum_{\underline{k}} \xi_{\underline{k}} \gamma_{i_1} \gamma_{i_2} \cdots \gamma_{i_k}
\end{equation}
where the sum over multiindex $\underline{k} = (i_1, i_2, \ldots, i_k)$ indicates the sum over $k$ non decreasing indices $1 \le {i_1} \le {i_2} \le \cdots \le {i_k} \le 2m$ and contains $\bino{2m}{k}$ terms. One easily shows \cite{BudinichP_1989} that the field coefficient is given by
$$
\xi_{\underline{k}} = \frac{1}{2^{m}} \langle B \varphi, \gamma^{i_k} \cdots \gamma^{i_2} \gamma^{i_1} \omega \rangle
$$
where $\gamma^i = (-1)^{i+1} \gamma_i$ so that $\frac{1}{2} \anticomm{\gamma^i}{\gamma_j} = \delta^{i}_{j}$.
%and at the end,
%\begin{equation}
%\label{multivector_expansion}
%\mu \cong \omega \otimes \varphi^* = \frac{1}{2^{m}} \sum_{k = 0}^{2 m} \sum_{\underline{k}} \langle B \varphi, \gamma^{i_k} \cdots \gamma^{i_2} \gamma^{i_1} \omega \rangle \gamma_{i_1} \gamma_{i_2} \cdots \gamma_{i_k} \dotinformula
%\end{equation}
Any $\mu \in \myCl{m}{m}{g}$ can also be expanded in the EFB, but we leave this for future research.
\opt{margin_notes}{\mynote{mbh.note: still missing spinorial expansion pp. 394.5 - 394.18; do we need it ?}}

The multivector expansion (\ref{multivector_expansion}) remains obviously valid whichever the basis of $V$, \eg replacing the $\gamma_i$ with the Witt basis (\ref{formula_Witt_basis}). To ease this passage we begin writing $\gamma_{i_1} \gamma_{i_2} \cdots \gamma_{i_k}$ in the Witt basis. Clearly it is enough to replace each $\gamma$ using (\ref{formula_Witt_basis}) but it is worth noting that each $\gamma$ appears in $\gamma_{i_1} \gamma_{i_2} \cdots \gamma_{i_k}$ either ``single'', \eg like $\gamma_{1}$ in $\gamma_{1}\gamma_{4} \cdots$, or ``married'' \ie in couples like $\gamma_{2 i - 1} \gamma_{2 i}$. With (\ref{formula_Witt_basis}) it is easily seen that each ``single'' $\gamma_i$ can be written as $p_i \pm q_i$ the sign depending on $i$ parity, whereas for each ``married'' couple we saw already that $\gamma_{2 i - 1} \gamma_{2 i} = \comm{q_i}{p_i}$ so that, at the end
\begin{equation}
\label{multivector_Witt_expansion}
\gamma_{i_1} \gamma_{i_2} \cdots \gamma_{i_k} = (p_{i_1} \pm q_{i_1}) \cdots (p_{i_l} \pm q_{i_l}) \comm{q_{j_1}}{p_{j_1}} \cdots \comm{q_{j_r}}{p_{j_r}}
\end{equation}
where we shifted all the commutators to the right since they commute with all other elements and where $l$ is the number of the singles and $r$ that of the couples and $l + 2 r = k$%
\footnote{We remark that each of the $2^l$ terms of the expansion of the single $\gamma$'s can come from $2^l$ different $\gamma$ multivectors since single $\gamma_i$ have either an even or an odd index \eg $p_1 p_2$ can come from $\gamma_1 \gamma_3, \gamma_1 \gamma_4, \gamma_2 \gamma_3$ or $\gamma_2 \gamma_4$. On the other hand, the commutators $\comm{q_{j}}{p_{j}}$ originate from just one $\gamma$ multivector, \ie each multivector determines uniquely all the married couples. It is simple to see that $k \pmod{2} \le l \le \min(k, 2 m - k)$ and $\max(0, k - m) \le r \le \lfloor \frac{k}{2}\rfloor$.}%
. Clearly in this form $\gamma_{i_1} \gamma_{i_2} \cdots \gamma_{i_k}$ expands in a sum of exactly $2^{l + r}$ terms, all of the same grade $k$.

We start proving a technical proposition that allows to calculate the field coefficients $\xi_{\underline{k}}$ of (\ref{multivector_expansion}) transformed in the Witt basis with (\ref{multivector_Witt_expansion}).

\begin{MS_Proposition}
\label{multivector_expansion_Witt_basis}
Let $x_i$ represent $q_i$ or $p_i$ and $y_j$ represent $q_j p_j$ or $p_j q_j$: the field coefficient of the term $x_{i_1} \cdots x_{i_l} y_{j_1} \cdots y_{j_r}$ of (\ref{multivector_expansion}) expressed in the Witt basis is given by:
\begin{displaymath}
\pm 2^{l+r-m} \langle B \varphi, \myconjugate{x}_{i_l} \cdots \myconjugate{x}_{i_1} y_{j_r} \cdots y_{j_1} \omega \rangle
\end{displaymath}
where $\myconjugate{x}_i = C x_i C^{-1}$ (\ref{C_def}) \ie $q_i = C p_i C^{-1}$ and viceversa.
\end{MS_Proposition}
\begin{proof}
With (\ref{multivector_Witt_expansion}) plugged in the multivector expansion (\ref{multivector_expansion}) one obtains
$$
\omega \otimes \varphi^* = \sum_{k = 0}^{2 m} \sum_{\underline{k}} \xi_{\underline{k}} (p_{i_1} \pm q_{i_1}) \cdots (p_{i_h} \pm q_{i_h}) \comm{q_{j_1}}{p_{j_1}} \cdots \comm{q_{j_s}}{p_{j_s}}
$$
and left multiplying both sides by $\myconjugate{x}_{i_l} \cdots \myconjugate{x}_{i_1} y_{j_r} \cdots y_{j_1}$ and taking the trace we have, for the left part of the equality,
$$
\mytrace{\myconjugate{x}_{i_l} \cdots \myconjugate{x}_{i_1} y_{j_r} \cdots y_{j_1} \omega \otimes \varphi^*} = \langle B \varphi, \myconjugate{x}_{i_l} \cdots \myconjugate{x}_{i_1} y_{j_r} \cdots y_{j_1} \omega \rangle \dotinformula
$$
Before calculating the result for the right part we remark that the term $\myconjugate{x}_{i_l} \cdots \myconjugate{x}_{i_1} y_{j_r} \cdots y_{j_1}$ by which we left multiplied can come \emph{only} from one of the $2^{l + r}$ terms of the expansion (\ref{multivector_Witt_expansion}) in which the $\gamma$ multivector had grade $l + 2 r$. Multiplying the right side of (\ref{multivector_expansion}) by any $\gamma$ multivector of grade $t$ and taking the trace, by the properties of the trace of $\gamma$ multivectors, one selects, in the sum over $k$ only the term with $k = t$ since all other terms have zero trace. By (\ref{multivector_Witt_expansion}) this holds also in our case and we can deduce that for our expansion of $\omega \otimes \varphi^*$ the first sum over $k$ disappears since terms with nonzero trace have necessarily $l + 2r = h + 2s = k$ and we obtain
\opt{margin_notes}{\mynote{mbh.note: this point is delicate, see details at p. 394.14'}}
$$
\sum_{\underline{k}} \xi_{\underline{k}} \mytrace{\myconjugate{x}_{i_l} \cdots \myconjugate{x}_{i_1} y_{j_r} \cdots y_{j_1} (p_{i_1} \pm q_{i_1}) \cdots (p_{i_h} \pm q_{i_h}) \comm{q_{j_1}}{p_{j_1}} \cdots \comm{q_{j_s}}{p_{j_s}}}
$$
and in calculating the product we remark that $\myconjugate{x}_{i} (p_i \pm q_i) = \pm \myconjugate{x}_{i} x_i$ and $y_j \comm{q_{j}}{p_{j}} = \pm y_j$. Moreover any trace containing in the product any isolated $p_i$, $q_i$ or $\comm{q_{j}}{p_{j}}$ is null. Thus the trace is not null if and only if $l = h$ and $r = s$ and each $\myconjugate{x}_{i}$ has its corresponding $(p_i \pm q_i)$ and each $y_j$ has its corresponding $\comm{q_{j}}{p_{j}}$. In summary we obtain
\begin{eqnarray*}
\mytrace{\myconjugate{x}_{i_l} \cdots \myconjugate{x}_{i_1} y_{j_r} \cdots y_{j_1} (p_{i_1} \pm q_{i_1}) \cdots (p_{i_l} \pm q_{i_l}) \comm{q_{j_1}}{p_{j_1}} \cdots \comm{q_{j_r}}{p_{j_r}}} = \\
%= \mytrace{\myconjugate{x}_{i_l} \cdots \myconjugate{x}_{i_1} (p_{i_1} \pm q_{i_1}) \cdots (p_{i_l} \pm q_{i_l}) y_{j_r} \cdots y_{j_1}} = \\
= \pm \mytrace{\myconjugate{x}_{i_l} x_{i_l} \cdots \myconjugate{x}_{i_1} x_{i_1} y_{j_r} \cdots y_{j_1}} = \pm 2^{m-l-r}
\end{eqnarray*}
and thus the thesis.
\end{proof}

With this result it is easy to give a simple proof to the following theorem due to Cartan \cite{Cartan_1937} and Chevalley \cite{Chevalley_1954} but we omit it in view of the fact that next theorem has a similar proof and that derives this one as a corollary.
\begin{MS_theorem}
\label{generalized_Cartan_theorem}
A nonzero spinor $\omega \in S$ is simple with $M(\omega) = \my_span{q_1, q_2, \ldots, q_{m}}$ if and only if it is a Weyl eigenvector (\ref{Weyl_eigenvector})
\opt{margin_notes}{\mynote{mbh.note: see prop.~8 p. 2130 of P\&T Phy\_434 \cite{BudinichP_1989} and prop.~5 p. 14 of T\&T Phy\_453 \cite{Trautman_Trautman_1994}}}
and the multivector expansion (\ref{multivector_expansion}) of $\omega \otimes \omega^*$ contains only the term $q_1 q_2 \cdots q_{m}$ \ie
$$
\omega \otimes \omega^* = \xi q_1 q_2 \cdots q_{m} \dotinformula
$$
\end{MS_theorem}

Up to now this has been the main theorem used to define a generic simple spinor and its application brings to the so called {\em constraint relations} explained in section~\ref{Physics_applications}. We now generalize this theorem relaxing the condition on $\omega \otimes \omega^*$ to a much milder one for $\omega \otimes \varphi^*$ \emph{ for any} $\varphi \in S$ that constitutes the main result of this work.
\begin{MS_theorem}
\label{generalized_Paolo_theorem}
A nonzero spinor $\omega \in S$ is simple with $M(\omega) = \my_span{q_1, q_2, \ldots, q_{m}}$ if and only if for any $\varphi \in S$
$$
\omega \otimes \varphi^* = \sum_{k = k_{m}}^{2 m} \sum_{\underline{k}} \xi_{\underline{k}} z_{i_1} z_{i_2} \cdots z_{i_k} \qquad z_{i} = q_{i}, q_{i} p_{i}
$$
where $k_{m} := \dim_\F M(\omega) \cap M(\varphi)$. Moreover it is sufficient to prove the relation for just one of the values $k_{m} \le k \le m$ to deduce that $\omega$ is simple.
\opt{margin_notes}{\mynote{mbh.note: at best of my knowledge neither Weyl nor $N(\varphi) > 0$ hypothesis are needed}}
\end{MS_theorem}
\begin{proof}
First of all we remark that there is no loss of generality in assuming $M(\omega) = \my_span{q_1, q_2, \ldots, q_{m}}$ since, by proposition~\ref{free_v1_vk}, we know that $q_1 q_2 \cdots q_{m} \propto v_1 v_2 \cdots v_{m}$ if $v_i$ span the same TNP and so it is easy to adapt the theorem to any maximal TNP in any basis.

We start proving a weaker version with $k_{m} \equiv 0$ for any $\varphi \in S$. Let's suppose first that $\omega$ is simple with $M(\omega) = \my_span{q_1, q_2, \ldots, q_{m}}$, for the field coefficients $\langle B \cdot , \cdot \rangle$ of the multivector expansion (\ref{multivector_expansion}) of $\omega \otimes \varphi^*$ we have, with $\gamma^i = (-1)^{i+1} \gamma_i$ and with (\ref{multivector_Witt_expansion}),
$$
\langle B \varphi, \gamma^{i_k} \cdots \gamma^{i_2} \gamma^{i_1} \omega \rangle = \pm \langle B \varphi, (p_{i_l} \pm q_{i_l}) \cdots (p_{i_1} \pm q_{i_1}) \comm{q_{j_r}}{p_{j_r}} \cdots \comm{q_{j_1}}{p_{j_1}} \omega \rangle
$$
and given the hypothesis on $\omega$ one easily sees that $\comm{q_{j}}{p_{j}} \omega = q_{j} p_{j} \omega$ and $(p_{i} \pm q_{i}) \omega = p_{i} \omega$ and so in the expansion of $\gamma^{i_k} \cdots \gamma^{i_2} \gamma^{i_1}$ in the Witt basis only one term out of the $2^{l + r}$ survives, namely:
$$
\langle B \varphi, \gamma^{i_k} \cdots \gamma^{i_2} \gamma^{i_1} \omega \rangle = \pm \langle B \varphi, p_{i_l} \cdots p_{i_1} \; q_{j_r} p_{j_r} \cdots q_{j_1} p_{j_1} \omega \rangle
$$
and with proposition~\ref{multivector_expansion_Witt_basis} we get the forward part of the theorem.

To prove the converse we remark that by proposition~\ref{multivector_expansion_Witt_basis} terms of the form $z_{i_1} z_{i_2} \cdots z_{i_k}$ have field coefficients $\langle B \varphi, p_{i_l} \cdots p_{i_1} \; q_{j_r} p_{j_r} \cdots q_{j_1} p_{j_1} \omega \rangle \ne 0$ while, by hypothesis, any term containing $p_i$ in the multivector expansion is zero, that implies that, for any $\varphi \in S$, $\langle B \varphi, q_i \omega \rangle = 0$. Since the inner product is not degenerate, $\langle B \varphi, q_i \omega \rangle = 0$ for any $\varphi \in S$ implies $q_i \omega = 0$ \ie $q_i \in M(\omega)$ since $\omega \ne 0$ by hypothesis. This procedure can be repeated for any $q_i$ giving $M(\omega) = \my_span{q_1, q_2, \ldots, q_{m}}$ \ie the thesis.

\opt{margin_notes}{\mynote{mbh.note: follows a slightly modified version of this paragraph (with respect to published one) due to the fact that old proposition~\ref{B_omega_varphi} was wrong, see a discussion at logbook pp. 526 ff.}}
We sharpen this result showing that the expansion of $\omega \otimes \varphi^*$ contains only terms with $k \ge k_{m}$: let $ \dim_\F M(\omega) \cap M(\varphi) = k_{m}$, \ie $\my_span{q_{i_1}, q_{i_2}, \ldots, q_{i_{k_m}}} \subseteq M(\varphi)$, by proposition~\ref{B_omega_varphi} $\langle B \varphi, p_{i_l} \cdots p_{i_2} p_{i_1} \omega \rangle = 0$ for all $l < k_{m}$ since at least $k_m$ $p_i$ must be present to ``shadow'' the $k_{m}$ $q_i$ that belong to $M(\omega) \cap M(\varphi)$ and thus, necessarily, that in the expansion, $k \ge k_{m}$.

We remark that the procedure can be restricted to any particular value of $k > 0$ and the proof remains valid. For example for $k = 1$ we can prove the theorem in $V$, deduce that $\omega$ is simple and derive the result for all other values of $k$.
\end{proof}

It is clear that choosing $\varphi = \omega$ then $k_{m} = m$ and we obtain as a corollary the theorem~\ref{generalized_Cartan_theorem} of Cartan and Chevalley; moreover the case $k=1$ of this theorem gives proposition~7 of \cite{BudinichP_1989}. Another difference between the two theorems is that here the hypothesis of the spinor $\omega$ being a Weyl eigenvector is not needed.

With this theorem one can prove that $\omega$ is simple with $M(\omega) = \my_span{q_1, q_2, \ldots, q_{m}}$ requiring that only $m$ constraints $\langle B \varphi, q_i \omega \rangle = 0 \quad i = 1, 2, \ldots, m$ are satisfied for any $\varphi \in S$.

\opt{margin_notes}{\mynote{mbh.note: see comments at start of section 2.6 of T\&T \cite{Trautman_Trautman_1994}}}
All results of this work are obtained in the hypothesis of even dimensional spaces $V = \F^{2 m}$ and $(m,m)$ signature for real spaces. This is customary in these studies because these cases are simpler to tackle given that the maximal TNP have dimension $m$. It looks quite plausible that, as for theorem~\ref{generalized_Cartan_theorem}, these results hold in a more general settings for any field of characteristics $\ne 2$, this being a direction for further investigations.

\section{Applications to Physics}
\label{Physics_applications}

In a seminal paper Berkovits \cite{Berkovits_2000_0} proposed a super-Poincaré covariant quantization of the superstring by means of simple (pure) spinors that uses Cartan Chevalley theorem~\ref{generalized_Cartan_theorem} as its starting point to define simple spinor $\omega$.

The idea is that to satisfy theorem~\ref{generalized_Cartan_theorem} and be simple a spinor $\omega$ must be a Weyl eigenvector (\ref{Weyl_eigenvector}) and all the terms of the multi vector expansion (\ref{multivector_expansion}) must be zero except one with $k = m$. By known results one can prove \cite{BudinichP_1989} that $\xi_{\underline{k}} = 2^{-m} B(\omega, \gamma^{i_k} \cdots \gamma^{i_2} \gamma^{i_1} \omega) = 0$ for $m - k \equiv 1,2,3 \pmod{4}$ so that to apply the theorem one needs to impose $B(\omega, \gamma^{i_k} \cdots \gamma^{i_2} \gamma^{i_1} \omega) = 0$ for just $m - k \equiv 0 \pmod{4}$ and for $k < m$ (for Hodge duality, see \cite{BudinichP_1989}). This implies a number of constraints of the order of $\bino{2m}{m-4}$ growing exponentially with $m$. For example for $m = 8$ one has to satisfy $1821$ constraints: one $B(\omega, \omega) = 0$ and $\bino{16}{4} = 1820$ constraints $B(\omega, \gamma^{i_4} \gamma^{i_3} \gamma^{i_2} \gamma^{i_1} \omega) = 0$.

In case of $2m = 10$-dimensional space of \cite{Berkovits_2000_0} one needs to satisfy just $\bino{10}{1} = 10$ constraints that in our formalism reads:
$$
B(\omega, \gamma^i \omega) = 0 \qquad i = 1, \ldots, 10 \dotinformula
$$
Five years later, in a subsequent paper \cite{Berkovits_Nekrasov_2005}, this simple spinor approach was extended to $11$ and $12$-dimensional space with $m = 6$ and the authors mention that they cannot attach physical interpretation to the $\bino{12}{2} = 66$ simple spinor constraints generated in this case:
$$
B(\omega, \gamma^{i_2} \gamma^{i_1} \omega) = 0 \qquad i_1, i_2 = 1, \ldots, 12 \dotinformula
$$
A possible road to explore could apply theorem~\ref{generalized_Paolo_theorem} and given the TNP $M(\omega) = \my_span{x_1, x_2, \ldots, x_{12}}$ one could replace these $66$ constraints with just $12$
$$
B(\omega, x_{i} \varphi) = 0 \qquad i = 1, \ldots, 12
$$
moreover for any $\varphi \in S$ and without the request of spinors being Weyl.

\smallskip

In a completely different field the results of this work apply to the recent proposal by Pavsic \cite{Pavsic_2010} that multiple spinor spaces $S_{h \circ g}$ can support mirror particles, see \eg \cite{Okun_2007}. As pointed out in section~\ref{Spinor_spaces}, \myCl{m}{m}{g}, as a vector space, is the direct sum of spinor subspaces of different $h \circ g-$signatures. Each of these $2^m$ spinor spaces carry faithful and irreducible representations of \myCl{m}{m}{g} and since the algebra is central simple they are isomorphic. One can take full advantage of the EFB formalism introduced in section~\ref{EFB_formalism} to easily derive that if $h \circ g'$ differs from $h \circ g$ in sites: $i_1, i_2, \ldots, i_k$ then
$$
S_{h \circ g'} = S_{h \circ g} (p_{i_1} + q_{i_1}) (p_{i_2} + q_{i_2}) \cdots (p_{i_k} + q_{i_k})
$$
and the new spinor has same chirality (\ref{Weyl_eigenvector}) but possibly different global parity (\ref{EFB_are_left_Weyl}) $\g_eigen' = (-1)^k \g_eigen$ and these matters deserve deeper investigations to fully evaluate this proposal.

%\newpage

\section{Conclusions}
We investigated the rich relations between null vectors and spinors exploiting some properties of the Extended Fock Basis.

With propositions~\ref{generic_multiple_v_spinor} and \ref{free_v1_vk} one can write explicitly the most general spinor corresponding to any vector subspace made entirely of null vectors.

We saw also that to define a generic simple spinor using the theorem of Cartan and Chevalley a number of constraint relations exponential in $m$ have to be satisfied. On the other hand, specifying the Totally Null Planes, \eg $\my_span{q_1, q_2, \ldots, q_{m}}$, then the definition of the corresponding simple spinor $\omega$ is straightforward with quoted propositions or with theorem~\ref{generalized_Paolo_theorem} that requires the satisfaction of only $m$ constraints.

This paper contains a first set of results obtained exploiting the EFB, some more are emerging and are due to come out in the near future. They will all make part of a program whose goal is to reinterpret the elements of geometry as made entirely of simple spinors \cite{Cartan_1937, BudinichP_2008} for which EFB seems particularly apt since with this basis it is possible to express very neatly all elements of \myCl{m}{m}{g}, scalars, vectors and multivectors, in terms of simple spinors.

\vspace{0.8 cm}

\subsection*{Dedication}
This paper is dedicated to the memory of my father Paolo Budinich who passed away in November 2013 not before transferring me his enthusiasm for simple spinors.

%\newpage

\vspace{0.8 cm}

\opt{x,std,AACA}{% in all cases - but arXiv, JMP & JOPA - standard BibTeX bibliography
\bibliographystyle{plain} % or plain or.... see e.g.\ http://amath.colorado.edu/documentation/LaTeX/reference/faq/bibstyles.html#styles

\bibliography{mbh}
}

\opt{arXiv,JMP,JOPA}{% only for arXiv, JMP & JOPA we need to include here the L•A•S•T version of file .bbl
}

\opt{final_notes}{
\newpage

\section{Things to do, notes, etc.......}

\noindent General things to do:
\begin{itemize}
\item simplicity and all that, logbook p. 394.21, Spinorial chessboard p. 10 (and 14).
\item Similarity and equivalence classes; logbook p. 394.28, 394.33.
\item Inner product $\langle B \cdot , \cdot \rangle$ stuff; logbook p. 481 (resumè), pp. 518, 520, 526.
\end{itemize}

\bigskip
.................
\smallskip

Unfortunately proposition~\ref{B_omega_varphi} was \emph{wrong!} a correct version has been replaced together with a slightly modified of but last paragraph of the proof of Theorem~\ref{generalized_Paolo_theorem}.

\bigskip
.................
\smallskip

This part was just before proposition~3 (in a wrong version, without ``stars'') and was cited nowhere else

$C$ (\ref{C_def}) defines an automorphism on the entire algebra and its action on $\mu \in \myCl{m}{m}{g}$ is given by $\myconjugate{\mu} = C \sum_{a} \xi_{a b}^\star \Psi_{a b} C^{-1}$ and one needs the transformation property of $\Psi_{a b}$. From (\ref{EFB_def}) and $C p_i C^{-1} = q_i$ one realizes that in the expression of $\Psi_{a b}$ all $p_i$ are replaced by $q_i$ and viceversa. This means that the $g$-signature remains unaltered whereas the $h$-signature is reversed, and so the $h \circ g$-signature is reversed as well. In summary this means that
\begin{equation*}
\label{C_transforms}
\myconjugate{\Psi}_{a b} = C \Psi_{a b} C^{-1} = \Psi_{-a -b}
\end{equation*}
\ie that the conjugation reverses both signatures so that for the generic element of the algebra one obtains
$$
\myconjugate{\mu} = \sum_{a b} \xi_{a b}^\star \Psi_{-a -b} = \sum_{a b} \xi_{-a -b}^\star \Psi_{a b}
$$
showing that, like for vectors, also for any algebra element one can choose to exchange the EFB elements \emph{or} the field coefficients $\xi_{a b}$.
\opt{margin_notes}{\mynote{mbh.note: eliminated every hint that this conjugation coincides with c.c. when $\F = \C$.}}

\bigskip
.................
\smallskip

This is the old \textbf{Appendix} replaced in the paper with a citation to \cite{Trautman_Trautman_1994}

%\newpage

%\subsection*{Appendix}
\begin{MS_Proposition}
\label{Phi_prop}
Given the generic spinor $\Phi$ with all $2^m$ terms of the expansion (\ref{Phi_def}) different from zero then for all nonzero $v \in V_0$ $v \Phi \ne 0$.
\opt{margin_notes}{\mynote{mbh.note: for a different proof of the first part see \cite{BudinichP_1989} (23) p. 2127}}
\end{MS_Proposition}
\begin{proof}
Given any nonzero $v \in V_0$ in proposition~\ref{S_bisect_prop} it is shown that the subspace $S_{\myconjugate{v}}$ is of dimension $2^{m - 1}$ so there exist many values of the coefficients $\xi_a$ of (\ref{Phi_def}) giving $v \Phi \ne 0$.

In the following example we show that the hypothesis is strict, \ie that if just one of the coefficients of the generic spinor (\ref{Phi_def}) is zero, \eg $\xi_a = 0$, then there exists $v \in V_0$ such that $v \Phi = 0$.

We proceed bottom up, \ie we start from $v \in V_0$ with which we build $\omega' := v \Phi \ne 0$ and, clearly, $v \omega' = v^2 \Phi = 0$. Then we show that $\omega'$ has necessarily at least one coefficient zero while all others can be nonzero.

Let's take $v = \sum_{i = 1}^m p_i$ that clearly is in $V_0$, then
$$
\omega' := v \Phi = \sum_{i = 1}^m p_i \Phi \ne 0
$$
and we start observing that for any $i$ the spinor $p_i \Phi$ does not contain the component $\Psi_0 = q_1 q_2 \cdots q_m$ since in $p_i \Phi$ the term $p_i \xi_0 \Psi_0$ gives
$$
p_i \xi_0 \Psi_0 = (-1)^{i - 1} \xi_0 q_1 q_2 \cdots p_i q_i \cdots q_m
$$
and it is also easy to show that $p_i \xi_a \Psi_a \ne \xi_a \Psi_0$ for any $a$ and thus we can conclude that the component $\Psi_0$ does not appear in $\omega'$.

We show now that all the other components of $\omega'$ can be nonzero and we start observing that each $p_i \Phi$ is made by a sum of terms that have all the $2^{m - 1}$ $h-$ signatures with $h_i = -1$, in other words in each $p_i \Phi$ only the $i$-th part of the $h$-signature is ``freezed'' at $h_i = -1$. But when we calculate the sum $\sum_{i = 1}^m p_i \Phi$ all $2^m$ $h$-signatures different from $\Psi_0$ will be represented. A given $h$ signature can appear in $1, 2, \ldots, m$ of the elements $p_i \Phi$; for example the signature $(-1,1,1,\ldots,1)$ comes only from $p_1 \Phi$ and in particular from $p_1 \xi_0 \Psi_0$ while $- e$ comes from all the $m$ terms $p_i \Phi$ and will thus have as field coefficient a sum of $m$ different coefficients $\xi$ of $\Phi$. We remark that each of the components has a sum of all different field coefficients, \eg if a given component $\Psi_a$ appears both in $p_i \Phi$ and in $p_j \Phi$ it will come from different components of $\Phi$ and thus the coefficients of $\Psi_a$ in $\omega'$ is a sum of different coefficients of $\Phi$. If we want $\omega' = v \Phi$ to have all components represented it is sufficient to choose the initial $2^m$ coefficients of $\Phi$ to take values $2^0, 2^1, \ldots, 2^{2^m - 1}$ because any signed sum of any subset of $1, 2, \ldots, m$ of these numbers can never be zero.
\end{proof}

\bigskip
.................
\smallskip

A nice formula showing that any simple spinor is Weyl....
$$
\Gamma v_1 \cdots v_m \Phi = \det{A} \Gamma q_1 \cdots q_m \Phi = \pm \det{A} \Phi
$$

\bigskip
.................
\smallskip

Old formula (2) now in correct form: see log. p. 505

The isomorphism (of vector spaces) $\myCl{m}{m}{g} \cong \Lambda V$ with the Grassmann algebra leads \cite{Chevalley_1954} to the following useful formula for the Clifford product $v \mu$ of any two elements $v \in V, \mu \in \myCl{m}{m}{g}$
\begin{equation}
\label{Clifford_product}
v \mu := v \JJ \mu + v \wedge \mu
\end{equation}
where $v \JJ \mu$ represents the {\em contraction} of $v$ with $\mu$ and $v \wedge \mu$ is the {\em exterior} or {\em wedge product}.

\bigskip
.................
\smallskip

\opt{margin_notes}{\mynote{mbh.note: This is the theorem of C \& C with its own demo.}}
With this proposition we can give a different proof to the following theorem due to Cartan \cite{Cartan_1937} and Chevalley \cite{Chevalley_1954}.
\begin{MS_theorem}
%\label{generalized_Cartan_theorem}
A nonzero spinor $\omega \in S$ is simple with $M(\omega) = \my_span{q_1, q_2, \ldots, q_{m}}$ if and only if the multivector expansion (\ref{multivector_expansion}) of $\omega \otimes \omega^*$ contains just the term $q_1 q_2 \cdots q_{m}$ \ie
$$
\omega \otimes \omega^* = \xi q_1 q_2 \cdots q_{m} \qquad \xi \in \F \dotinformula
$$
\end{MS_theorem}
\begin{proof}
First of all we remark that there is no loss of generality in assuming $M(\omega) = \my_span{q_1, q_2, \ldots, q_{m}}$ since, by proposition~\ref{free_v1_vk}, we know that $q_1 q_2 \cdots q_{m} \propto v_1 v_2 \cdots v_{m}$ if $v_i$ span the same TNP and so it is easy to adapt the theorem to any TNP in any basis.

So let's suppose first that $\omega$ is simple with $M(\omega) = \my_span{q_1, q_2, \ldots, q_{m}}$, then in expansion (\ref{multivector_expansion}) with (\ref{multivector_Witt_expansion}) the field coefficients are given by $\langle B \omega, \gamma^{i_k} \cdots \gamma^{i_1} \omega \rangle = \langle B \omega, (p_{i_1} \pm q_{i_1}) \cdots (p_{i_l} \pm q_{i_l}) \comm{q_{j_1}}{p_{j_1}} \cdots \comm{q_{j_r}}{p_{j_r}} \omega \rangle$ and with the hypothesis and proposition~\ref{B_omega_varphi} one gets easily that the only nonzero term can be $\langle B \omega, p_m \cdots p_2 p_1 \omega \rangle$ that, with proposition~\ref{multivector_expansion_Witt_basis}, proves the the forward part of the theorem.

To prove the converse let's suppose
$$
\omega \otimes \omega^* = \xi q_1 q_2 \cdots q_{m}
$$
that implies that all terms but one of multivector expansion (\ref{multivector_expansion}) are zero and by proposition~\ref{multivector_expansion_Witt_basis} we know that the field coefficient is:
$$
0 \ne \xi = \langle B \omega, p_m \cdots p_2 p_1 \omega \rangle
$$
from which we can deduce $p_m \cdots p_2 p_1 \omega \ne 0$ and thus $M(\omega) \cap \my_span{p_1, p_2, \ldots, p_{m}} = \{0\}$. From $ 0 = \langle B \omega, q_m p_{m-1} \cdots p_2 p_1 \omega \rangle = \langle B q_m \omega, p_{m-1} \cdots p_2 p_1 \omega \rangle$ and with proposition~\ref{B_omega_varphi} it follows $\dim_\F M(q_m \omega) \cap M(p_{m-1} \cdots p_2 p_1 \omega) > 0$. Since we know that $M(\omega) \cap \my_span{p_1, p_2, \ldots, p_{m}} = \{0\}$ it follows that also $M(q_m \omega) \cap \my_span{p_1, p_2, \ldots, p_{m-1}} = \{0\}$ and thus the direction in common between $q_m \omega$ and $p_{m-1} \cdots p_2 p_1 \omega$ is not in $\my_span{p_1, p_2, \ldots, p_{m}}$. It cannot be $p_m$ since $p_m p_{m-1} \cdots p_2 p_1 \omega \ne 0$ and for the very same reason it cannot be $q_i$ for $i = 1, \ldots, m-1$. It follows that necessarily $q_m p_{m-1} \cdots p_2 p_1 \omega = 0$ and thus $q_m \omega = 0$. Since this procedure works not only for $q_m$ but for any $q_i$ we get $M(\omega) = \my_span{q_1, q_2, \ldots, q_{m}}$ and thus the thesis.
\end{proof}

\bigskip
.................
\smallskip

This is the old subsection ``Totally Null Planes again''

%\subsection{Totally Null Planes again}
%\label{TNPs}
We prove here some technical propositions needed in the sequel.
\opt{margin_notes}{\mynote{mbh.note: see log-book pg. 484' \& 484.1.}}
\begin{MS_Proposition}
\label{TNPs_prop_basis}
Given $k \le m$ nonzero vectors $v_i \in V$, they form a TNP if and only if $\anticomm{v_i}{v_j} = 0$ for any $1 \le i, j \le k$.
\end{MS_Proposition}
\begin{proof}
Since the vectors have to be all null this implies $v_i^2 = 0 = \anticomm{v_i}{v_i}$ and since any of their linear combinations have to be as well null this implies $\left(\sum_{i = 1}^k \alpha_i v_i\right)^2 = \sum_{i > j} \alpha_i \alpha_j \anticomm{v_i}{v_j} = 0$ and since this must hold for any $\alpha_i$ this implies $\anticomm{v_i}{v_j} = 0$ for any $i, j$. The converse is immediate.
\end{proof}
\opt{margin_notes}{\mynote{mbh.note: see log-book p. 505}}
It follows that for two vectors forming a TNP $v_1 v_2 = v_1 \wedge v_2$. This easily generalizes for $k$ vectors to $v_1 v_2 \cdots v_k = v_1 \wedge v_2 \wedge \cdots \wedge v_k$ that, for the property of the external product, proves:
\begin{MS_Corollary}
\label{TNPs_prop_dim}
Given $k \le m$ nonzero vectors $v_i \in V$ forming a TNP, the dimension of the TNP is $k$ if and only if $v_1 v_2 \cdots v_k \ne 0$.
\end{MS_Corollary}
\begin{MS_Proposition}
\label{TNPs_+1_prop}
Given $k-1 < m$ nonzero vectors that form a TNP of dimension $k-1$, a $k$-th null vector can be added to them to form a TNP of dimension $k$ if and only if neither $v_{k}$ nor $\myconjugate{v}_{k}$ are in $\my_span{ v_1, v_2, \ldots, v_{k-1}}$.
\end{MS_Proposition}
\begin{proof}
That necessarily $v_{k} \notin \my_span{ v_1, v_2, \ldots, v_{k-1}}$ is fairly obvious; to show that also $\myconjugate{v}_{k}$ must not be in $\my_span{ v_1, v_2, \ldots, v_{k-1}}$ let's suppose the contrary, \ie $\myconjugate{v}_{k} = \sum_{i = 1}^{k-1} \alpha_i v_i$, then since $0 \ne \anticomm{v_{k}}{\myconjugate{v}_{k}} = \sum_{i = 1}^{k-1} \alpha_i \anticomm{v_{k}}{v_{i}}$ and thus $\anticomm{v_{k}}{v_{i}} \ne 0$ for some $i$ and thus $v_{k}$ cannot form a TNP with $v_1 v_2 \cdots v_{k-1}$. The converse is immediate.
\end{proof}

\bigskip
.................
\smallskip

This is the old Proposition~17....

\begin{MS_Proposition}
%\label{multivector_expansion_Witt_basis}
Given a basis of $V$ such that $\anticomm{v_i}{v_j} = 0$ for $i \ne j$ then, in the multivector expansion (\ref{multivector_expansion}) of $\omega \otimes \varphi^*$, the field coefficient of the term $v_{i_1} v_{i_2} \cdots v_{i_k}$, for which $\anticomm{v_{i_j}}{v_{i_l}} = 0$ for any ${i_j}, {i_l}$ and $1 \le {i_1} \le {i_2} \le \cdots \le {i_k} \le m$, is given by $2^{k-m} \langle B \varphi, \myconjugate{v}_{i_k} \cdots \myconjugate{v}_{i_2} \myconjugate{v}_{i_1} \omega \rangle$ where $\myconjugate{v}_i$ is the conjugate of $v_i$.
\opt{margin_notes}{\mynote{mbh.prob: double check this!}}
\end{MS_Proposition}
\begin{proof}
We start observing that, by (\ref{formula_Witt_basis}), $\gamma^i = (-1)^{i+1} \gamma_i = \myconjugate{\gamma}_{i}$ so that the proposition is already proved for a non-null basis of $V$ and we just need to prove the proposition when $v_i$ form a null basis of $V$. In this case we may assume, without loss of generality,
$$
\anticomm{v_{i}}{v_{j}} = \anticomm{\myconjugate{v}_{i}}{\myconjugate{v}_{j}} = 0
\qquad
\anticomm{v_{i}}{\myconjugate{v}_{j}} = \delta_{i j}
\qquad i, j = 1, \ldots, m
$$
and one easily finds that $\mytrace{v_{i}} = \mytrace{\myconjugate{v}_{j}} = 0$ and that $\mytrace{v_{i} \myconjugate{v}_{i}} = 2^{m - 1}$. To calculate the value of the coefficients we calculate $\mytrace{x_{i_k} \cdots x_{i_2} x_{i_1} \omega \otimes \varphi^*}$ and since $\mytrace{x_{i_k} \cdots x_{i_2} x_{i_1} v_{i_1} v_{i_2} \cdots v_{i_l}}$ is different from zero if and only if $k = l$ and $x_i = \myconjugate{v}_{i}$ for all $i = 1, \ldots, k$ and in this case $\mytrace{\myconjugate{v}_{i_k} \cdots \myconjugate{v}_{i_2} \myconjugate{v}_{i_1} v_{i_1} v_{i_2} \cdots v_{i_k}}= 2^{m - k}$, we are done.
\end{proof}

\bigskip
.................
\smallskip

In EFB the generic Witt basis vector $x_k \in \{ p_k, q_k \}$ can be written concisely using the basic property (\ref{formula_Witt_basis_properties}), $\anticomm{q_j}{p_j} = \Identity$, as
$$
x_k = \anticomm{q_1}{p_1} \anticomm{q_2}{p_2} \cdots \anticomm{q_{k-1}}{p_{k-1}} \; x_k \; \anticomm{q_{k+1}}{p_{k+1}} \cdots \anticomm{q_m}{p_m}
$$
that, in expanded form, is a sum of $2^{m - 1}$ EFB elements; for a generic $v \in V$ the property extends to all terms of the expansion $v = \sum\limits_{i=1}^{m} \left( \alpha_{i} p_{i} + \beta_{i} q_{i} \right)$.

As a consequence in EFB the vectors are scattered through many $h-$ and $g-$signatures and it is thus useful to characterize them:
\begin{MS_Proposition}
$\phi \in \myCl{m}{m}{g} - \F$ is a vector if and only if $\anticomm{\phi}{v} \in \F$ for any vector $v$ of \myCl{m}{m}{g}
\end{MS_Proposition}
We just proved one part of the proposition, to prove the other we suppose $\anticomm{\phi}{v} \in \F$ for any vector $v$ and we start taking $v = x_k \in \{ p_k, q_k \}$. Given the hypothesis and the EFB expansion
$$
\phi = \sum\limits_{i=1}^{2^{2 m}} \gamma_{i} \Omega_{i} \qquad \Omega_{i} = \omega_1 \omega_2 \cdots \omega_m
$$
necessarily all EFB elements $\Omega_{i}$ must contain the $k$-th component $\omega_k$ and it is a simple exercise to prove that necessarily $\omega_k \in \my_span{p_k q_k - q_k p_k, p_k, q_k}$.

Let's see the consequences on other $\omega$ components: suppose $\omega_k = q_k$ since $\anticomm{q_k}{p_k} = 1$ it is clear that all other components $\omega_i$ with $i \ne k$ must be $1$ otherwise $\anticomm{\Omega}{p_k} \notin \F$. The same observation applies to the case $\omega_k = p_k$.
%
%On the contrary all components $\ne k$ of $\omega_1 \omega_2 \cdots \omega_{k-1} p_k \omega_{k+1} \cdots \omega_{m}$ could take any value because in this case the anticommutator is always null, but our hypothesis is $\anticomm{\omega}{v} \in \F$ for all v and so if we take $v = q_k$ we easily find that also in this case $\omega_1 = \omega_2 = \cdots = \omega_{k-1} = \omega_{k+1} = \cdots = \omega_{m} = 1$.
%
We are thus left with the third possibility $\omega_k = p_k q_k - q_k p_k$; we take now $v = p_l$ with $l \ne k$ we see that while $\anticomm{p_k}{p_l} = \anticomm{q_k}{p_l} = 0 \in \F$ in this case $\anticomm{\Omega}{p_l} \notin \F$ so we must conclude that $\Omega \in \my_span{p_k, q_k}$. Given the arbitrary choice of the initial vector $p_k$ we can conclude that in general $\Omega \in \my_span{p_1, q_1, \ldots, p_m, q_m}$. \qed

\bigskip
.................
\smallskip

We note that the product of a Witt basis vector $x_k \in \{ p_k, q_k \}$ by an EFB element $\omega$ is
$$
x_k \omega = \left\{
\begin{array}{l l}
0 & \quad \mbox{if $h(x_k) = h_k(\omega)$} \\
\omega' \ne 0 & \quad \mbox{if $h(x_k) = - h_k(\omega)$}
\end{array} \right.
$$
and thus $(p_k + q_k) \omega = \omega'$ and
$$
h_k(\omega') = - h_k(\omega) = h(x_k) \qquad g_k(\omega') = - g_k(\omega)
$$
that shows that left multiplication by unitary vector $p_k + q_k$ has the effect of inverting both $h-$ and $g-$signatures of the corresponding components of $\omega$. On the other hand
$$
\omega x_k = \left\{
\begin{array}{l l}
0 & \quad \mbox{if $h(x_k) = h_k(\omega)$} \\
\omega' \ne 0 & \quad \mbox{if $h(x_k) = - h_k(\omega)$}
\end{array} \right.
$$
and thus $\omega (p_k + q_k) = \omega'$ with
$$
h_k(\omega') = h_k(\omega) = - h(x_k) \qquad g_k(\omega') = - g_k(\omega)
$$
that shows that right multiplication by $p_k + q_k$ has the effect of maintaining $h-$ while inverting $g-$signature of the corresponding component of $\omega$. So it is simple to prove:
\begin{MS_Proposition}
Given any $\omega, \phi$ EFB elements of \myCl{m}{m}{g} then, given $u_i, v_i \in \{p_i + q_i, p_i q_i + q_i p_i\}$, there is one and only one transformation such that
$$
\phi = u_1 u_2 \cdots u_m \; \omega \; v_1 v_2 \cdots v_m \dotinformula
$$
Each $u_i, v_i$ has a well defined $g-$signature; respectively $-1$ and $+1$, that identifies it uniquely. The $g-$signatures of $u_i$ and $v_i$ elements are
\begin{align*}
g(u_i) & = h_i(\omega) h_i(\phi) \\
g(v_i) & = h_i(\omega) h_i(\phi) g_i(\omega) g_i(\phi)
\end{align*}
\end{MS_Proposition}
We consider first $\omega' = u_1 u_2 \cdots u_m \, \omega$ given what we have just seen any $u_i$ with $g(u_i) = 1$ is $u_i = p_i q_i + q_i p_i = 1$ and thus has no effect on the corresponding $\omega_i$, on the contrary when $g(u_i) = -1$ and $u_i = p_i + q_i$ its effect is to change sign to both $h_i(\omega)$ and $g_i(\omega)$. Given the prescription $g(u_i) = h_i(\omega) h_i(\phi)$ we can deduce that only when $h_i(\omega) \ne h_i(\phi)$ the corresponding $h-$ and $g-$signatures of $\omega$ are changed so that we can deduce $h(\omega') = h(\phi)$ and $g(\omega') = h(\omega) h(\phi) g(\omega)$. Considering now $\omega' \, v_1 v_2 \cdots v_m$ and $g(v_i) = h_i(\omega) h_i(\phi) g_i(\omega) g_i(\phi) = g_i(\omega') g_i(\phi)$ we see that $g(v_i) = -1$ only when $g_i(\omega') \ne g_i(\phi)$ and thus $g(\omega' \, v_1 v_2 \cdots v_m) = g(\phi)$ and given that the $h-$signature is not changed by right multiplication of $v_i$ and that $h-$ and $g-$signatures identify uniquely an EFB element the proposition is proved. \qed

We remember here that the Pin group consists of products of unit vectors and Spin is its subgroup consisting of products of even sequences of such vectors.

These properties are the base of the
\begin{MS_Proposition}
Let $\omega$ be a simple spinors of \myCl{m}{m}{g} then, given a vector $v$ such that $v^2 \ne 0$, $v \omega$ is simple, of opposite chirality and with $M(v \omega) = v M(\omega) v$ while $\omega v$ is simple, of same chirality and with $M(\omega v) = M(\omega)$.
\end{MS_Proposition}
It is simple to see that $\Gamma p_i = - p_i \Gamma$ and $\Gamma q_i = - q_i \Gamma$ and thus that for any vector $\Gamma v = - v \Gamma$ from which easily descend that $v \omega$ is a Weyl eigenvector of opposite chirality of $\omega$. For any vector $u \in M(\omega)$ we easily have $u \omega = 0 = u v v \omega = v u v v \omega$. Since $v u v$ has an inverse given by $v^{-1} = \frac{v}{v^2}$ it defines a proper transformation and thus the transformed TNP $v M(\omega) v$ is of maximal dimension and the spinor $v \omega$ is thus simple. For the other case we see that obviously $\omega v$ is also a Weyl eigenvector of same chirality and that for any $u \in M(\omega)$ $u \omega v = 0$. \qed

} % note finali: stampate solo se all'inizio c'è l'opzione final_notes

\end{document}